\documentclass[a4paper,11pt]{article}
\usepackage{longtable}
\usepackage{titlesec}
\usepackage{xcolor}
\usepackage{booktabs}
\usepackage{ragged2e}
\usepackage[utf8]{inputenc}
\usepackage{bm}
\usepackage{helvet}
\usepackage[T1]{fontenc}
\usepackage{lmodern}
\usepackage[a4paper,margin=0.75in]{geometry}
\usepackage{amsmath,amssymb,amsfonts}
\usepackage{graphicx}
\usepackage{booktabs}
\usepackage{array}
\usepackage{titlesec}
\titleclass{\subsubsubsection}{straight}[
\subsubsection]
\newcounter{subsubsubsection}[subsubsection]
\renewcommand\thesubsubsubsection{\thesubsubsection.\arabic{subsubsubsection}}
\titleformat{\subsubsubsection}[runin]
  {\normalfont\normalsize\bfseries}{\thesubsubsubsection}{1em}{}
\titlespacing*{\subsubsubsection}{0pt}{3.25ex plus 1ex minus .2ex}{1em}

\usepackage{titlesec}
\usepackage{tabularx}
\usepackage{longtable}
\usepackage{float}
\usepackage{caption}
\usepackage{enumitem}
\usepackage[hidelinks]{hyperref}
\usepackage{authblk}
\usepackage{float}
\usepackage{algorithm}
\usepackage{algpseudocode}
\renewcommand{\arraystretch}{1.2}

\title{
\begin{center}

\vspace{0.9cm}
{\bf \LARGE
E-UDE: A Scientific Machine Learning Model for Malware Propagation Dynamics}
\vspace{0.9cm}

\end{center}
}

\author[1]{Vineet Kumar}
\author[2,*]{Samiran Ghosh}

\author[2]{V Anil Kumar}
\vspace{.3cm}
\affil[1]{School of Information Technology,
Rajiv Gandhi Proudyogiki Vishwavidyalaya (RGPV),
Bhopal, 462033, India}
\vspace{0.5cm}
\affil[2]{Bigdata Research and Supercomputing Division,
CSIR National Institute of Data Science and AI,
Wind Tunnel Road, Bengaluru, 560037, India}

\affil[*]{
Corresponding author: \texttt{samiran.4pi@csir.res.in}}

\date{}

\begin{document}

\maketitle

\begin{abstract}

Understanding the malware propagation dynamics over the Internet is crucial for effectively controlling its spread. Existing mechanistic and data-driven models in the context of malware propagation dynamics, have some issues with either interpretability or predictive accuracy. The Universal Differential Equation framework (UDE) is a hybrid framework that combines both the mechanistic and data-driven approaches. This study presents an Enhanced Universal Differential Equation (E-UDE) framework that explicitly incorporates more detailed key factors that influence malware propagation, including transmission dynamics, network topology, congestion effects, external attacks, defensive responses, and security degradation. The neural residual part of this framework also allows the model to capture nonlinear effects that are not adequately represented by the underlying differential equations. Evaluation on real-world malware propagation data shows that the E-UDE achieves the best overall performance, attaining an overall RMSE of 783.70, an MAE of 421.65, and a MAPE of 13.36\%. Compared with the Classical ODE and the UDE models, the proposed approach reduces RMSE by approximately 66\% and 39\%, respectively. Its robustness to predictive capability is further evaluated using a 25-75 chronological train–test split. In this experiment, the E-UDE achieves a Test RMSE of 634.6, a Test MAE of 419.3, and a Test Correlation of 0.964. The E-UDE maintains relatively stable prediction performance as the noise level increases, showing limited degradation compared with the baseline models. Furthermore, this E-UDE model enables us to perform several scenario analyses under different cybersecurity scenarios, providing a useful framework for better understanding the effective control factors. To study the wide applicability of the E-UDE framework, the model performance is assessed by two different datasets, namely, Code Red dataset and Witty Worm dataset.

\end{abstract}


\vspace{2em}
\section{Introduction}
Modern society has become increasingly dependent on the digital world through multiple devices connected to the Internet, which has indeed improved services in many sectors such as healthcare, finance, transportation, and critical infrastructure. On the other hand, threats to the cyber world have also evolved over the years. Over the past few decades, several malware attacks have been witnessed that have disrupted digital infrastructure and eventually impacted human society. The Code Red worm outbreak in 2001 rapidly compromised hundreds of thousands of internet hosts very rapidly \cite{moore2002code,moore2003internet}. The financial loss associated with the Code Red epidemic and its subsequent strains was estimated to be approximately \$2.6 billion \cite{moore2003internet}. The SQL Slammer worm in 2003 \cite{moore2003inside}, the Conficker worm in 2008 \cite{shin2011large,zhang2015hybrid}, the WannaCry ransomware outbreak in 2017 \cite{trautman2019wannacry}, among others, have significantly impacted Internet traffic.

To better understand and control the self-propagation mechanism of malware spread in computer networks, several attempts have been made in the existing literature using SIR-type epidemic models \cite{cheng2010modeling,wei2019modeling,al2019analysis}. More detailed models incorporating several phenomena, including different stages of infectiousness and various intervention strategies, have also been studied \cite{peng2025dynamic,wang2017sadi,aleja2022compartmental,godoi2023spatio}. In recent years, researchers have increasingly integrated machine learning techniques with compartmental epidemic models to improve the understanding and prediction of malware propagation \cite{severt2023comparison}. Deep learning and time-series forecasting techniques have been extensively applied to model and predict cyber attack rates. Commonly used approaches include ARIMA, GARCH, hybrid ARIMA–GARCH models, Recurrent Neural Networks (RNNs), Long Short-Term Memory (LSTM) networks, and Bidirectional RNN–LSTM (BRNN-LSTM) architectures \cite{fang2019deep,sun2023cyber,peng2017modeling}. Recent advances in physics-informed machine learning have demonstrated that integrating mechanistic knowledge with deep neural networks can substantially improve predictive performance while maintaining a degree of model interpretability \cite{ngoy2021combining,sulaiman2024defense}.

Hybrid modeling is well suited to complex dynamical systems in which some governing mechanisms are understood, whereas other processes remain uncertain or difficult to formulate explicitly. Universal Differential Equation (UDE) is a mathematical framework connecting the physical laws with the machine learning approaches \cite{rackauckas2020universal}. In this context, \cite{pappu2025understanding} developed a Universal Differential Equation (UDE) framework for malware propagation that integrates mechanistic differential equations with neural-network-based components. The authors have shown that the UDE model gives better performance compared to the ODE model and the Neural ODE model \cite{pappu2025understanding}. However, the UDE model in that study takes into account the generic mechanisms of malware propagation process, including logistic growth, external forcing, and suppression etc. This generic representation may not capture the detailed multi-state specific infection progression. The important states such as susceptibility, infection, response, and congestion, remain implicit in the model, rather than being modeled as separate dynamic compartments. This generic representation restricts the model's ability to investigate stage-specific interventions, quantify the influence of individual parameters, and analysis of several cybersecurity scenarios. To overcome these limitations, we introduce an Enhanced Universal Differential Equation (E-UDE) framework that incorporates multiple infection related compartments along with the neural network as a residual component. This compartmental structure gives a more detailed representation of malware propagation. The main contributions of this study are summarized as follows:

\begin{itemize}
\item We develop an Enhanced Universal Differential Equation (E-UDE) model that extends the existing UDE by explicitly considering four malware propagation states: susceptible, infected, response, and congestion.


\item We demonstrate that the proposed E-UDE model provides improved and robust predictive performance compared with existing ODE, Neural ODE, and UDE models.

\item We show that the compartmental structure of the E-UDE enables detailed parameter-driven scenario analysis under different cybersecurity scenarios, allowing the effectiveness and cost-efficiency of different cyber defense strategies to be systematically evaluated.

\item Furthermore, we establish that our proposed E-UDE model is effective across different types of propagation datasets.
\end{itemize}

Overall, the novelty of the proposed E-UDE framework lies in combining the detailed compartmental malware propagation model with the neural residual learning within a unified modeling framework. By explicitly representing distinct propagation and defense-related states, the proposed approach provides greater mechanistic interpretability and enables state-specific cybersecurity scenario analysis that is not readily possible with existing UDE formulation for malware propagation. This makes the framework relevant for evaluating targeted cybersecurity interventions under diverse operational scenarios.

The rest of the paper is organized as follows. In Section~\ref{data_description}, we describe the data used in this work. In Section~\ref{model_survey}, we describe some of the existing models for malware propagation dynamics relevant to this research. We then introduce the proposed E-UDE methodology in Section~\ref{proposed_methodology}. In Section~\ref{results}, we present the results of the proposed model using real malware propagation data, followed by a robustness analysis of the proposed model in Section~\ref{robustness}. The E-UDE-based scenario analysis is performed in Section~\ref{scenarios}. The general applicability of the E-UDE is then investigated using different datasets in Section~\ref{cross_data}. Finally, the discussion and conclusion of the work are presented in Section~\ref{discussion}.

\section{Data description}\label{data_description}

This study employs the publicly available Code Red worm dataset provided by CAIDA \cite{caida_codered_dataset}, which contains darknet telescope observations of scanning activities generated during the Code Red worm outbreak. In addition to the Code Red dataset, we use the CAIDA UCSD Dataset on the
Witty Worm to evaluate the proposed model on a different malware outbreak.
The Witty worm dataset was collected using the UCSD Network Telescope
during the Witty worm outbreak from 19-24 March, 2004
\cite{caida_witty}. The dataset contains information about Witty-infected
hosts and their observed activity, including infection start and end times,
activity duration, traffic volume, and other propagation-related information.
These observations provide a real-world view of how the Witty worm spread
across the Internet. To transform the raw event-based records into a continuous-time signal suitable for differential equation modeling, a three-stage preprocessing pipeline was applied (as used in \cite{pappu2025understanding}): (a) Temporal aggregation; (b) Smoothing; and (c) Continuous-time interpolation.

\begin{figure}[ht!]
    \centering
    \includegraphics[width=0.8\linewidth]{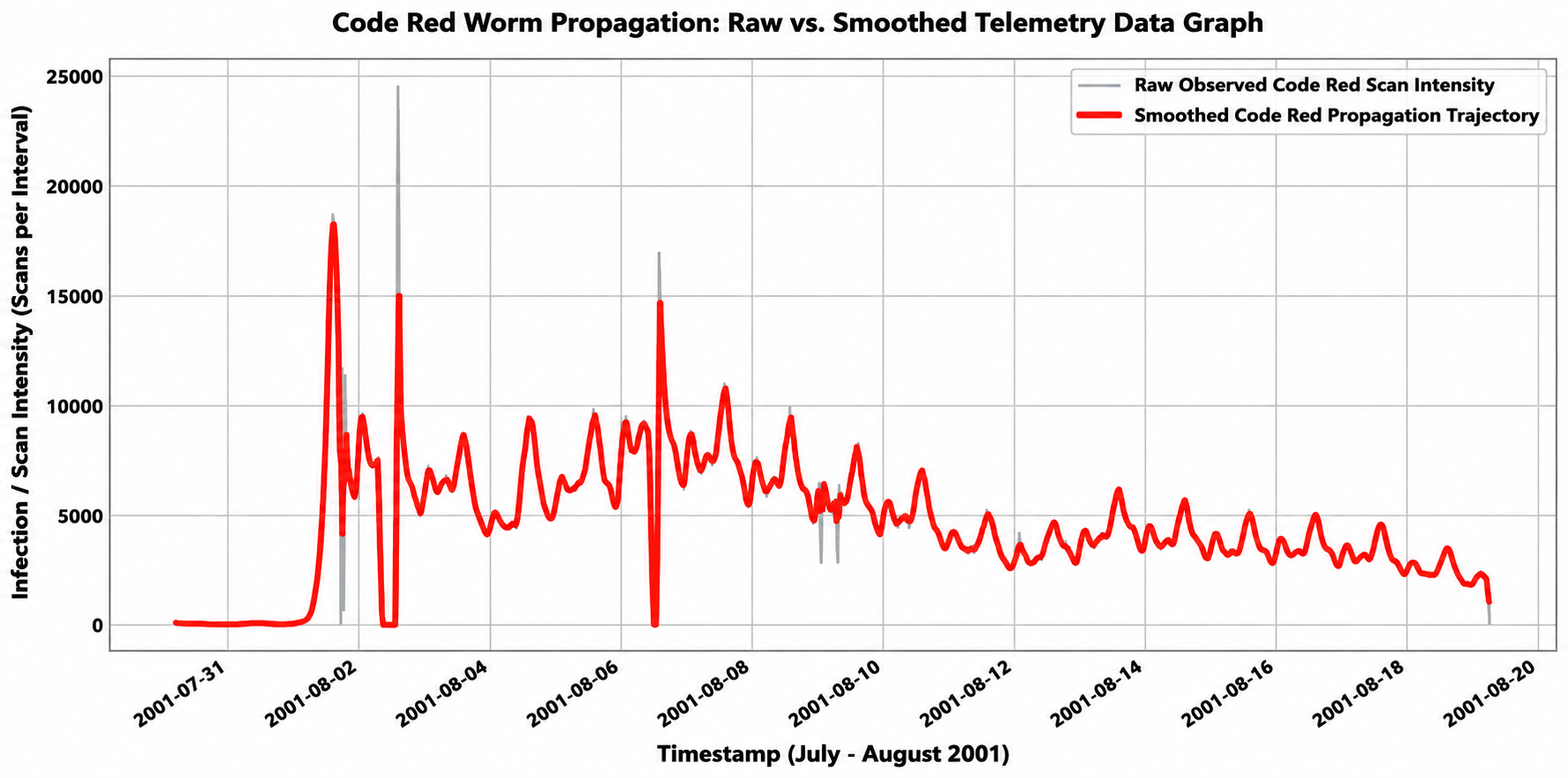}
    \caption{Code Red worm scan intensity showing raw telemetry and 3-point smoothed trajectory at 30-minute intervals (time step $\Delta t = 30\,\text{min}$, and number of total sample points $N = 963$).}
    \label{fig:code_red_smooth_data}
\end{figure}

Through temporal aggregation, smoothing, and interpolation, the raw scan records are transformed into a continuous infection intensity function that captures the dynamics of the outbreak while providing a stable input for Scientific Machine Learning (SciML) and differential equation-based modeling. The resulting infection intensity function denoted by $\eta(t)$ can be evaluated at any time during numerical integration and serves as the external forcing signal in the proposed malware propagation models. See more details about preprocessing in the Appendix A and reference \cite{pappu2025understanding}.

\section{Malware propagation models}\label{model_survey}
In this section, we discuss four existing modelling approaches for malware propagation dynamics.

\paragraph{ODE model.}

The classical epidemic model is the most widely used framework to describe the epidemic progression, which was introduced to described human epidemic. However, later the malware propagation dynamics are also described by such models.

The foundation of mathematical models to describe the transmission dynamics of biological epidemics was developed by D. Bernoulli in the 18th century and by W. O. Kermack and A. G. McKendrick in the early 20th century \cite{kermack1927contribution,kermack1932contributions,kermack1933contributions}. 
At the end of the last century, Cohen and Murray proposed that malware propagation in the cyber world could be analyzed using biological epidemic models \cite{cohen1987computer,murray1988application}.

The one dimensional classical ODE model to describe the malware propagation taking into account the important factors e.g., logistic growth, external forcing, suppression effect, and adaptive feedback, is given by (see \cite{pappu2025understanding,zou2002code,fang2022statistical}): 

\begin{equation}\label{eqn_1}
\frac{dM}{dt}
=
\alpha(t)M\left(1-\frac{M}{K}\right)
+\eta(t)
-\beta M^2
+\kappa M\log(1+M),
\end{equation}
where $M(t)$ represents the malware infection intensity at time $t$. Here, $\alpha(t)$ is the time dependent growth rate, $K$ is the carrying capacity, $\beta$ is the supression coefficient, $\kappa$ is the feedback strength, and $\eta(t)$ is the external forcing. More details can be found in \cite{pappu2025understanding}.

\paragraph{Neural ODE model.}

Neural ordinary differential equation (ODE) model is a differential equation where the rate of change is determined  by a neural network instead of some physics based rules and the neural network gets trained from the real data \cite{pappu2025understanding}. The Neural ODE can be represented as follows:
\begin{equation}
\frac{dM}{dt}=N_{\psi}(M,t),
\end{equation}
where $M$ is the malware infection intensity and $N_{\psi}$ is a neural network with trainable parameters $\psi$. The detailed architecture and implementation algorithm are given in \cite{pappu2025understanding}.

The Neural ODE has the advantage of learning complex nonlinear relationships directly from the available data, without requiring the underlying dynamics to be specified in advance through manually constructed equations. This framework does not inherently incorporate domain-specific knowledge from cybersecurity or epidemiological modeling. Consequently, obtaining reliable and stable estimates may depend more heavily on the availability of sufficiently large and representative training datasets. Without physical or mechanistic constraints, the model may also learn unrealistic behaviors that fit the training data well but generalize poorly to unseen malware propagation scenarios.

\paragraph{Universal Differential Equation (UDE) model.}

The Universal Differential Equation (UDE) is a hybrid type of model that takes into account both the physics based mechanism and the neural network architecture \cite{pappu2025understanding}. The UDE model can be defined as follows \cite{pappu2025understanding}: 

\begin{equation}
\frac{dM}{dt}
=
\alpha(t)M\left(1-\frac{M}{K}\right)
+\eta(t)
-\beta M^2
+N_{\phi}(M),
\end{equation}
where
$
\alpha(t)=\alpha_0e^{-p_{\mathrm{decay}}t/t_{\max}},
$
and $N_{\phi}(M)$ is the neural network feedback term.

The UDE is a hybrid model that combines a traditional malware propagation equation with a neural network. Instead of learning everything from data, the network focuses only on the unknown feedback mechanisms. Through symbolic recovery, the authors in \cite{pappu2025understanding} found that real-world malware propagation is strongly shaped by suppression effects such as network saturation and defensive actions.

\section{Proposed framework: Enhanced Universal Differential Equation (E-UDE) model}\label{proposed_methodology}

\begin{figure}[ht!]
    \centering
    \includegraphics[width=1\linewidth]{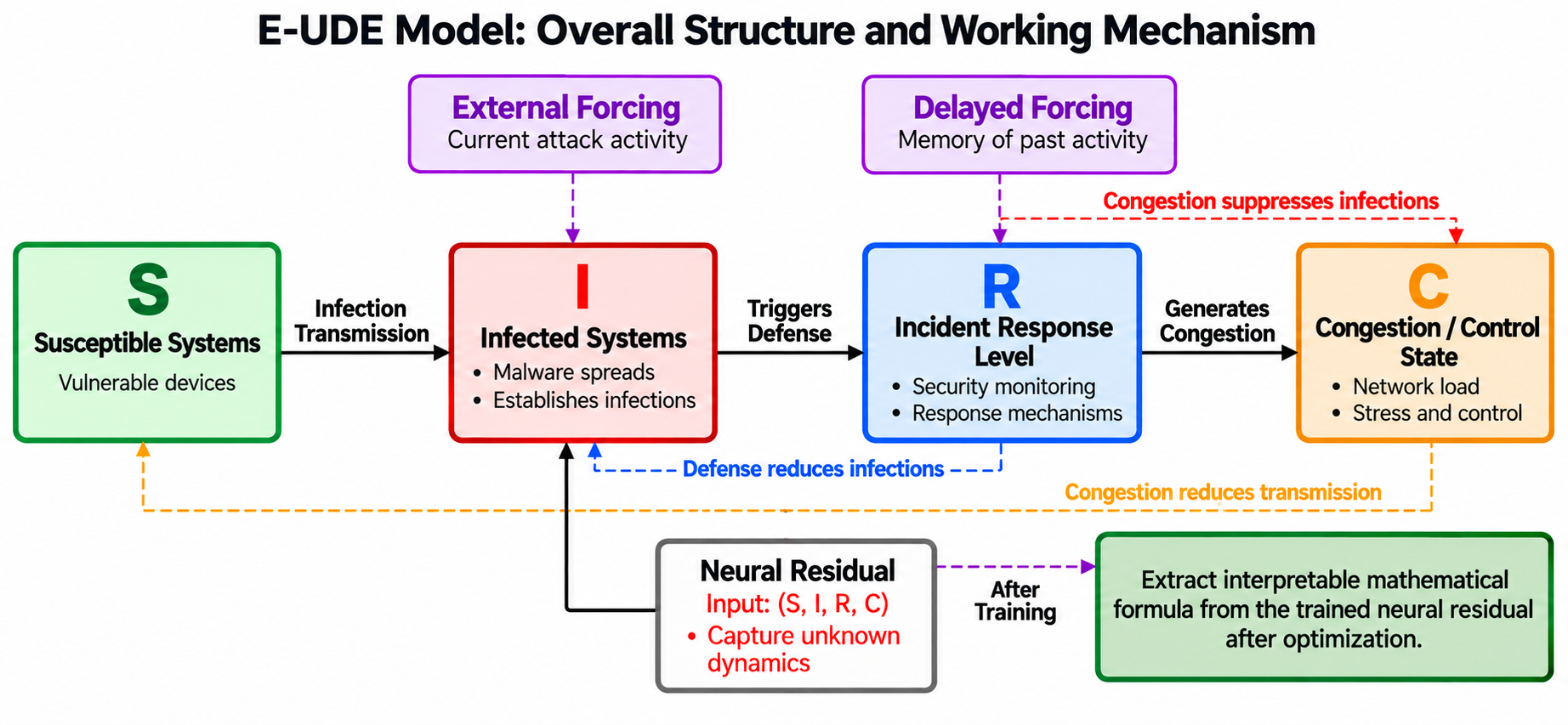}
        \caption{Schematic diagram of the E-UDE model.}
    \label{fig:model_flow}
\end{figure}

Among the three models considered, the ODE model, the Neural ODE model, and the Universal Differential Equation (UDE) model, the UDE model demonstrated the best performance in terms of both data fitting and predictive accuracy \cite{pappu2025understanding}. However, the existing UDE model incorporates only a limited set of physics-based mechanisms, such as logistic growth, external forcing, and suppression effects. It does not explicitly account for the different operational states within the network, including susceptible, infected, incident response, and congestion states. Consequently, its ability to perform detailed parameter-driven scenario analysis and to evaluate the impact of stage-specific control interventions is limited. To address these limitations, we propose an Enhanced UDE (E-UDE) model that explicitly incorporates these different stages of infection and network dynamics. The proposed malware propagation model is developed as a four-compartment Universal Differential Equation (UDE) system consisting of Susceptible $(S(t))$, Infected $(I(t))$, Incident Response $(R(t))$, and Congestion $(C(t))$ compartments. The first equation corresponding to $S$ describes how healthy devices become exposed to malware and tracks their return to susceptibility following recovery; the second equation corresponding to $I$ models the spread and mitigation of the infection, the third equation corresponding to $R$ captures the activation and decay of the network's automated defense and incident response systems, and the fourth equation corresponding to $C$ tracks the network congestion dynamics resulting from malicious activity and defensive traffic. The compartment-wise explanation is given below and the schematic diagram is given in Figure~\ref{fig:model_flow}.

\paragraph{Susceptible Compartment}

The susceptible equation describes the temporal evolution of vulnerable devices within a network. These systems may include desktop computers, servers, cloud instances, virtual machines, mobile devices, and Internet of Things (IoT) devices that have not yet been compromised but remain exposed to malware because of software vulnerabilities, configuration weaknesses, delayed security updates, or previously unknown zero-day exploits. As malware propagates throughout the network, susceptible devices are progressively infected and transferred to the infected compartment.
 The transmission mechanisms from susceptible to infected compartment include adaptive transmission dynamics, communication congestion, enterprise network topology, external attack sources, higher-order nonlinear cyber interactions, and incomplete remediation of infected devices.

The governing equation of the susceptible compartment including all these processes, is given by

\begin{equation}
\frac{dS}{dt}
=
-
\left[
\beta(t)
\left(\frac{I}{N}\right)
\exp\left(
-\alpha\frac{C}{M_{\max}}
\right)
+
\frac{F(t)}{N}
+
w_{\mathrm{res}}\phi_{\mathrm{trans}}\left(\frac{S}{N},\frac{I}{N},\frac{R}{N},\frac{C}{N}\right)
\right]S
+\gamma I.
\label{eq:susceptible}
\end{equation}

This equation integrates several mechanisms governing malware transmission in enterprise networks. The effective transmission rate is modeled as
$
\beta(t)=\beta_0\exp\left(-\frac{p_{\mathrm{decay}}t}{t_{\max}}\right),
$
where $\beta_0$ denotes the initial transmission rate, $p_{\mathrm{decay}}$ controls the variation of transmission rate, $t$ is the current time, and $t_{\max}$ is the maximum simulation time used to normalize the decay. Internal malware propagation follows the classical mass-action interaction term $SI/N$, where $S$, $I$, and $N$ denote the numbers of susceptible devices, infected devices, and the total network population, respectively. To account for network congestion, the infection process is modulated by the damping factor
$
\exp\left(-\alpha\frac{C}{M_{\max}}\right),
$
where $C$ represents the current congestion level, $M_{\max}$ is the maximum congestion capacity of the network, and $\alpha$ quantifies the sensitivity of malware transmission to congestion. Continuous infection attempts originating from external adversaries are incorporated through the normalized forcing term $F(t)/N$, where $F(t)$ denotes the time-dependent intensity of external attacks. Finally, unknown nonlinear interactions that cannot be explicitly modeled are represented by the residual term
$
w_{\mathrm{res}}\phi_{\mathrm{trans}}\left(\frac{S}{N},\frac{I}{N},\frac{R}{N},\frac{C}{N}\right),
$
where $\phi_{\mathrm{trans}}(\cdot)$ is a neural-network-based residual function that learns unmodeled transmission dynamics from data using the normalized system states $(S/N,I/N,R/N,C/N)$ as inputs, with $R$ denoting the incident response compartment. To ensure numerical stability, the input state is first bounded to physically meaningful values and normalized by the total network population $N$. Here, $w_{\mathrm{res}}$ denotes the learned residual weight that controls the contribution of the neural residual to the transmission dynamics. Furthermore, $\gamma$ represents the rate of incomplete remediation, accounting for infected devices that have had malware removed but remain unpatched, causing them to immediately transition back to the susceptible state.

 \paragraph{Infected Compartment}

The infected compartment represents devices that have been successfully compromised and are currently involved in malware propagation across the network. Once compromised, these devices can perform activities such as vulnerability scanning, exploit execution, lateral movement, phishing, command-and-control communication, and payload delivery. Such activities can lead to further infections among susceptible devices. The dynamics of the infected compartment are described by

\begin{equation}
\frac{dI}{dt} = \left[ \beta(t) \frac{I}{N} \exp\left(-\alpha \frac{C}{M_{\max}}\right) + \frac{F(t)}{N} + w_{\mathrm{res}}\phi_{\mathrm{trans}}\left(\frac{S}{N},\frac{I}{N},\frac{R}{N},\frac{C}{N}\right) \right] S - (\gamma + \delta R + \mu C) I.
\label{eq:infected}
\end{equation}

The equation accounts for both the addition of newly compromised devices and the removal of infected devices through different mitigation mechanisms. The inflow into the infected compartment is the same transmission term used in the susceptible equation,

$$
\left[ \beta(t) \frac{I}{N} \exp\left(-\alpha \frac{C}{M_{\max}}\right) + \frac{F(t)}{N} + w_{\mathrm{res}}\phi_{\mathrm{trans}}\left(\frac{S}{N},\frac{I}{N},\frac{R}{N},\frac{C}{N}\right) \right] S.
$$

The loss of infected devices occurs through three mechanisms. The first term, $\gamma I$, represents remediation that removes the malware but does not necessarily address the vulnerabilities that allowed the infection to occur. Here, $\gamma$ denotes the rate of such remediation, after which the affected devices can return to the susceptible compartment. The second term, $\delta R I$, represents active defense and containment. The parameter $\delta$ measures the effectiveness of the response process, while $R$ represents the level of active incident response. Their interaction reflects the increased removal of infected devices when defensive resources are actively engaged, for example through endpoint detection and response systems or firewall-based containment.

The final term, $\mu C I$, represents suppression caused by network congestion. The parameter $\mu$ controls the extent to which increasing congestion reduces the effective activity of infected devices and, consequently, their ability to continue participating in malware propagation.

\paragraph{Response Compartment}

The response compartment $R(t)$ represents the level of active cybersecurity response within the enterprise network. It reflects the extent to which security teams and automated defense mechanisms are engaged in detecting, containing, and responding to ongoing malware activity. Thus, $R(t)$ is used as a dynamic measure of the network's current defensive state.
The dynamics of the response compartment are given by

\begin{equation}
\frac{dR}{dt}
= \rho I-\omega R
+f_{\mathrm{def}}F(t).
\label{eq:response}
\end{equation}

The response level increases when the number of infected devices grows, as represented by the term $\rho I$. Here, $\rho$ denotes the strength of defensive measures. $\omega$ denotes the decay rate of defensive responses. $F(t)$ describes the intensity of external attack activity and $f_{\mathrm{def}}$ is the effectiveness of external forcing on defensive responses.







\paragraph{Congestion Compartment}

The congestion compartment denotes the level of network congestion developed during a malware outbreak. It captures the additional communication load generated by infected devices as they scan for vulnerable hosts, attempt exploits, communicate with command-and-control servers, send repeated connection requests, and carry out other malicious activities. These activities consume network bandwidth and processing capacity, which are both limited in an enterprise environment. As the malware spreads, the increase in malicious traffic can place greater pressure on the available network resources. The resulting congestion can, in turn, affect the ability of infected devices to communicate and propagate the malware effectively. The congestion compartment therefore acts as an intermediate state that links the malware activity with the condition of the underlying network infrastructure.
The dynamics of the congestion compartment are governed by

\begin{equation}
\frac{dC}{dt}
=
\sigma I
-
\tau C,
\label{eq:congestion}
\end{equation}

where $\sigma$ denotes the average traffic generation rate per infected device and $\tau$ is the congestion dissipation rate. This term models the ability of the enterprise network infrastructure to process and remove communication traffic through routing, packet forwarding, bandwidth allocation, timeout mechanisms, and packet loss. Since the dissipation rate is proportional to the current congestion level, it prevents unlimited accumulation of malicious traffic and ensures stable congestion dynamics.

The respective algorithm for the mechanistic part of E-UDE is given in Algorithm~\ref{alg:eude_compartments}.

\begin{algorithm}[ht!]
\small
\caption{Forward Integration of the 4-Compartment E-UDE System}
\label{alg:eude_compartments}
\begin{algorithmic}[1]

\Require Initial states $S_0, I_0, R_0, C_0$; total hosts $N$; capacity scale $M_{\max}$; physical parameters $\beta_0, p_{\mathrm{decay}}, t_{\max}, \alpha, \gamma, \delta, \mu, \rho, \omega, \sigma, \tau, g_{\mathrm{def}}$; memory weights $w_0, w_1, w_2$; neural parameters $w_{\mathrm{res}}, B_{\mathrm{res}}, \psi$; total steps $K$; step size $\Delta t$.

\Ensure Compartment trajectories $S[0:K], I[0:K], R[0:K], C[0:K]$

\State $S[0]\gets S_0,\ I[0]\gets I_0,\ R[0]\gets R_0,\ C[0]\gets C_0$

\For{$k=1$ \textbf{to} $K$}

    \State $t \gets k \Delta t$

    \State $\beta(t) \gets \beta_0 \exp\left(-p_{\mathrm{decay}} \frac{t}{t_{\max}}\right)$
    \Comment{Adaptive contact rate decay}

    \State $\eta_{\mathrm{eff}}(t) \gets w_0\eta(t) + w_1\eta(t-1) + w_2\eta(t-2)$
    \Comment{Delayed external attack forcing}

    \State $\phi_{\mathrm{cong}} \gets \exp\left(-\alpha \frac{C[k-1]}{M_{\max}}\right)$
    \Comment{Congestion-induced transmission damping}

    \State $\tilde{\mathbf{x}} \gets \left[ \frac{S[k-1]}{N},\ \frac{I[k-1]}{N},\ \frac{R[k-1]}{N},\ \frac{C[k-1]}{N} \right]$
    \Comment{Population-normalized state vector}

    \State $r_{\mathrm{raw}} \gets \mathrm{MLP}_{\psi}(\tilde{\mathbf{x}})$

    \State $\phi_{\mathrm{trans}} \gets B_{\mathrm{res}} \tanh\left(\frac{r_{\mathrm{raw}}}{B_{\mathrm{res}}}\right)$
    \Comment{Bounded neural residual}

    \State $\Phi[k] \gets \left[ \beta(t) \frac{I[k-1]}{N} \phi_{\mathrm{cong}} + \frac{\eta_{\mathrm{eff}}(t)}{N} + w_{\mathrm{res}} \phi_{\mathrm{trans}} \right] S[k-1]$
    \Comment{Conserved infection inflow}

    \State $F_{\mathrm{EDR}} \gets \delta R[k-1] I[k-1]$
    \Comment{Incident response / EDR containment}

    \State $F_{\mathrm{cong}} \gets \mu C[k-1] I[k-1]$
    \Comment{Congestion packet-drop suppression}

    \State $F_{\mathrm{rec}} \gets \gamma I[k-1]$
    \Comment{Incomplete remediation return}

    \State $\dot{S} \gets -\Phi[k] + F_{\mathrm{rec}}$
    \State $\dot{I} \gets \Phi[k] - F_{\mathrm{rec}} - F_{\mathrm{EDR}} - F_{\mathrm{cong}}$
    \State $\dot{R} \gets \rho I[k-1] - \omega R[k-1] + g_{\mathrm{def}} \eta_{\mathrm{eff}}(t)$
    \State $\dot{C} \gets \sigma I[k-1] - \tau C[k-1]$

    \State $S[k] \gets S[k-1] + \dot{S} \Delta t$
    \State $I[k] \gets I[k-1] + \dot{I} \Delta t$
    \State $R[k] \gets R[k-1] + \dot{R} \Delta t$
    \State $C[k] \gets C[k-1] + \dot{C} \Delta t$

    \State $S[k] \gets \operatorname{clamp}(S[k], 0, N)$
    \State $I[k] \gets \operatorname{clamp}(I[k], 0, N)$
    \State $R[k] \gets \max(0, R[k])$
    \State $C[k] \gets \max(0, C[k])$
    \Comment{Physical state non-negativity projection}

\EndFor

\State \Return $S, I, R, C$

\end{algorithmic}
\end{algorithm}

    
    

\subsubsection*{Neural Network Architecture and Residual Learning}

The proposed Enhanced Universal Differential Equation (E-UDE) augments the mechanistic epidemiological model with a lightweight neural network that learns only the unknown nonlinear transmission dynamics while preserving the interpretability of the governing equations. The neural residual is implemented as a fully connected Multi-Layer Perceptron (MLP) with a single hidden layer and hyperbolic tangent activation:

\begin{equation}
\begin{aligned}
r_{\mathrm{raw}} &= W_{2} \tanh\!\left( W_{1}\mathbf{x}+b_{1} \right) + b_{2}, \\[6pt]
\mathbf{x} &= [S/N, I/N, R/N, C/N]^T \in \mathbb{R}^{4}, \\
W_{1} &\in \mathbb{R}^{16 \times 4}, \quad b_{1} \in \mathbb{R}^{16}, \\
W_{2} &\in \mathbb{R}^{1 \times 16}, \quad b_{2} \in \mathbb{R}^{1},
\end{aligned}
\label{eq:mlp}
\end{equation}
where $\mathbf{x}$ is the normalized system state comprising the susceptible, infected, incident response, and congestion compartments. Here, $W_{1}$ and $W_{2}$ denote the weight matrices, while $b_{1}$ and $b_{2}$ are the corresponding bias vectors. 

The MLP consists of four input neurons, one hidden layer with sixteen neurons, and a single output neuron, resulting in $64$ input weights, $16$ hidden biases, $16$ output weights, and one output bias, for a total of $97$ trainable parameters, collectively denoted as $\theta_{\mathrm{residual}}$. Owing to its shallow architecture, the network remains computationally efficient while reducing the risk of overfitting.

To ensure numerical stability, the input state is first bounded to physically meaningful values and normalized by the total network population $N$. The normalized state is then propagated through the MLP to obtain the raw residual $r_{\mathrm{raw}}$, which is bounded using

\begin{equation}
r_{\mathrm{bounded}}
=
B_{\mathrm{res}}
\tanh\!\left(
\frac{r_{\mathrm{raw}}}{B_{\mathrm{res}}}
\right),
\label{eq:tanh_bound}
\end{equation}

where $B_{\mathrm{res}}$ is the learnable residual bound, initialized to $3.0$. The bounded residual is incorporated into the governing equations through the learnable scaling parameter

\begin{equation}
R_{\mathrm{neural}}
=
w_{\mathrm{res}}\,r_{\mathrm{bounded}},
\label{eq:scaled_residual}
\end{equation}

where $w_{\mathrm{res}}$ controls the contribution of the neural correction. The scaling parameter $w_{\mathrm{res}}$ is initialized to $0.5$. Both $w_{\mathrm{res}}$ and the residual bound $B_{\mathrm{res}}$ are optimized jointly with the remaining model parameters.

To preserve the dominance of the mechanistic model during the early optimization stages, the output layer is initialized with weights sampled uniformly from $[-10^{-3},10^{-3}]$ and zero biases. Consequently, the initial neural residual satisfies $R_{\mathrm{neural}}\approx0$, allowing the E-UDE to behave initially as the mechanistic model. As optimization proceeds, the neural network gradually learns only the residual nonlinear dynamics that are not captured by the physics-based equations, thereby improving predictive accuracy while maintaining numerical stability and physical interpretability. The neural residual computation algorithm is given in Algorithm~\ref{alg:eude_neural_residual}.

\begin{algorithm}[ht!]
\small
\caption{Neural Residual Computation in the E-UDE Framework}
\label{alg:eude_neural_residual}
\begin{algorithmic}[1]

\Require System state vector $x = [S, I, R, C]^T$
\Require Total population $N$, Numerical state clip bound $S_{\text{clip}}$
\Require MLP hidden layer parameters $W_1 \in \mathbb{R}^{16 \times 4}$, $b_1 \in \mathbb{R}^{16}$
\Require MLP output layer parameters $W_2 \in \mathbb{R}^{1 \times 16}$, $b_2 \in \mathbb{R}^{1}$
\Require Learnable residual bound $B_{\text{res}}$, Learnable scaling weight $w_{\text{res}}$
\Ensure Scaled neural transmission rate $r_{\phi}$

\State $x_{\text{bounded}} \gets \max(0, \min(x, S_{\text{clip}}))$ 
\Comment{Ensure numerical stability and non-negative populations}

\State $x_{\text{norm}} \gets \frac{x_{\text{bounded}}}{N}$
\Comment{Normalize inputs to $[0, 1]$ scale for the MLP}

\State $h \gets \tanh(W_1 x_{\text{norm}} + b_1)$
\Comment{Compute hidden layer activations (16 neurons)}

\State $r_{\text{raw}} \gets W_2 h + b_2$
\Comment{Compute linear output projection (1 neuron)}

\State $r_{\text{bounded}} \gets B_{\text{res}} \tanh\left(\frac{r_{\text{raw}}}{B_{\text{res}}}\right)$
\Comment{Enforce stability bound to prevent explosive gradients}

\State $r_{\phi} \gets w_{\text{res}} \cdot r_{\text{bounded}}$
\Comment{Scale by the learnable residual contribution weight}

\State \Return $r_{\phi}$

\end{algorithmic}
\end{algorithm}

\section{Results}\label{results}

In this section, we evaluate the predictive performance of the proposed Enhanced Universal Differential Equation (E-UDE) framework and compare it with existing differential equation and neural network based models.

\paragraph{Comparison of Model Predictions.}

\begin{figure}[ht!]
    \centering
\includegraphics[width=0.95\linewidth]{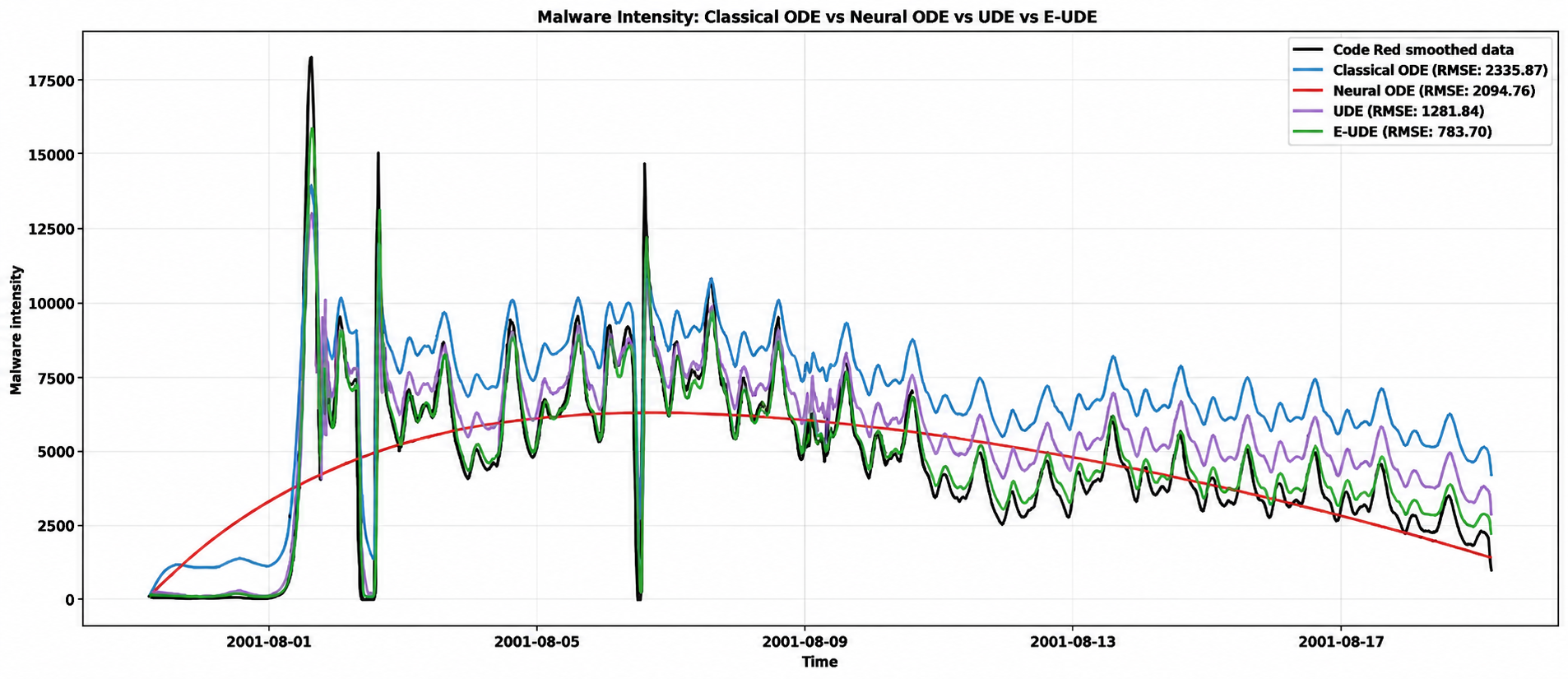}
    \caption{Comparison of different model predictions with Code Red worm propagation data.}
    \label{fig:fitting_data}
\end{figure}

\begin{figure}[ht!]
    \centering
    \includegraphics[height=8cm,width=0.8\linewidth]{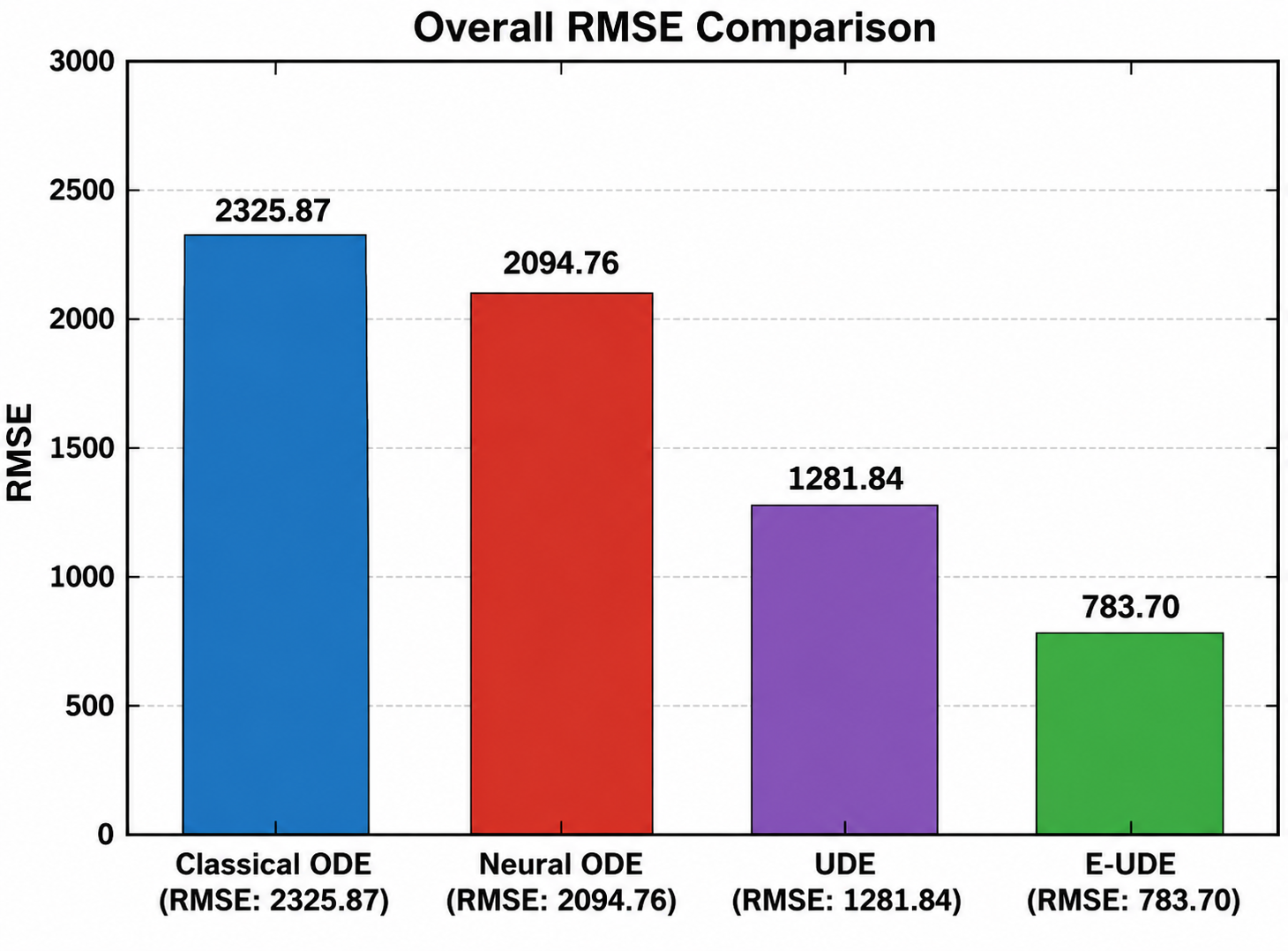}
    \caption{Overall RMSE comparison among different models.}
    \label{fig:rmse_comparison}
\end{figure}

Figure~\ref{fig:fitting_data} compares the malware propagation trajectories predicted by the different models. The choice of initial values and the estimated parameter values, together with the hyperparameter settings used in the E-UDE model, are summarized in Tables~\ref{tab:eude_parameters} and~\ref{tab:eude_training}, respectively. The corresponding overall RMSE comparison among different models is shown in Figure~\ref{fig:rmse_comparison}. The Classical ODE captures the overall outbreak trend but fails to reproduce several rapid variations observed in the data because of its limited nonlinear representation. The Neural ODE learns complex temporal dynamics directly from data but exhibits unstable predictions in certain phases of the outbreak. The UDE framework benefits from retaining known mechanistic relationships while allowing a neural component to learn the dynamics that are not adequately captured by the analytical formulation. In the proposed E-UDE, this combination results in the strongest correspondence with the measured malware-propagation trajectory over the complete study interval. The model successfully tracks the principal outbreak maxima as well as the smaller variations occurring between them, indicating that the compartment-based formulation can represent both large-scale and transient changes in the observed dynamics. The key structural differences between the conventional UDE formulation and the proposed E-UDE are summarized in Table~\ref{tab:ude_comparison}.


\begin{table*}[ht!]
\centering
\footnotesize
\renewcommand{\arraystretch}{1.15}

\caption{Initial and optimized parameters of the E-UDE model.}
\label{tab:eude_parameters}

\begin{minipage}{0.48\textwidth}
\centering
\begin{tabular}{lcc}
\toprule
\textbf{Parameter [Units]} & \textbf{Initial} & \textbf{Optimized} \\
\midrule
$\beta_0$ $[\text{h}^{-1}]$ & 0.0030 & 0.00257584 \\
$\gamma$ $[\text{h}^{-1}]$ & 1.2000 & 1.35530758 \\
$\delta$ $[\text{hosts}^{-1}\,\text{h}^{-1}]$ & $5.0\times10^{-8}$ & $5.91477\times10^{-8}$ \\
$\mu$ $[\text{hosts}^{-1}\,\text{h}^{-1}]$ & $1.0\times10^{-6}$ & $1.19116\times10^{-6}$ \\
$\rho$ $[\text{h}^{-1}]$ & 1.2000 & 1.34585500 \\
$\omega$ $[\text{h}^{-1}]$ & 0.1200 & 0.12605256 \\
$\sigma$ $[\text{h}^{-1}]$ & 0.0001 & 0.00011330 \\
$\tau$ $[\text{h}^{-1}]$ & 0.0800 & 0.06761630 \\

\bottomrule
\end{tabular}
\end{minipage}
\hfill
\begin{minipage}{0.48\textwidth}
\centering
\begin{tabular}{lcc}
\toprule
\textbf{Parameter [Units]} & \textbf{Initial} & \textbf{Optimized} \\
\midrule
$N$ $[\text{hosts}]$ & 91440 & 91440 \\
$p_{\rm decay}$ $[-]$ & 1.0000 & 1.14763343 \\
$\alpha$ $[-]$ & 0.1000 & 0.09459651 \\
$w_1$ $[-]$ & 0.6000 & 0.72449762 \\
$w_2$ $[-]$ & 0.3000 & 0.20678492 \\
$w_3$ $[-]$ & 0.1000 & 0.06871741 \\
$w_{\rm res}$ $[-]$ & 0.5000 & 0.70234162 \\
$B_{\rm res}$ $[-]$ & 3.0000 & 3.00100000 \\
$f_{\rm def}$ $[-]$ & 0.0100 & 0.01269044 \\
\bottomrule
\end{tabular}
\end{minipage}
\end{table*}


\begingroup

\renewcommand{\arraystretch}{1.25}

\begin{table}[ht!]
\centering
\small
\caption{Training and numerical configuration of the E-UDE model.}
\label{tab:eude_training}

\begin{tabularx}{\textwidth}{
>{\RaggedRight\arraybackslash}p{3.5cm}
>{\centering\arraybackslash}p{3.0cm}
>{\RaggedRight\arraybackslash}X
}

\toprule
\textbf{Hyper Parameter} &
\textbf{Value/Setting} &
\textbf{Description} \\
\midrule

Optimizer &
Adam &
Gradient-based optimizer used for joint training of the mechanistic,
neural, memory, and symbolic components. \\

Learning Rate &
0.001 &
Learning rate employed by the Adam optimizer for updating trainable
parameters. \\

Weight Decay &
$1\times10^{-5}$ &
$L_{2}$ regularization coefficient applied to neural parameters to reduce
overfitting. \\

Epochs &
300 &
Maximum number of training epochs used during E-UDE optimization. \\

ODE Solver &
Euler &
Numerical integration scheme employed to approximate the temporal evolution
of the E-UDE state equations. \\

ODE Tolerances &
$10^{-3}$ (atol/rtol) &
Absolute and relative numerical tolerance values specified in the model
configuration. \\

Gradient Clip &
1.0 &
Maximum gradient norm used to prevent excessively large parameter updates
during backpropagation through the dynamical system. \\

Early Stopping &
100 epochs &
Training patience interval used when the validation loss fails to improve
by at least $10^{-4}$. \\

Neural Residual Penalty &
0.01 &
Regularization coefficient applied to the magnitude of
$\mathcal{N}_{\phi}$ to discourage excessive dependence on the learned neural correction. \\

\bottomrule

\end{tabularx}
\end{table}

\endgroup


\begingroup
\color{blue}
\renewcommand{\arraystretch}{1.25}

\begin{longtable}{
>{\centering\arraybackslash}p{3.0cm}
>{\centering\arraybackslash}p{3.0cm}
>{\centering\arraybackslash}p{4.0cm}
>{\RaggedRight\arraybackslash}p{5.0cm}
}

\bottomrule
\endfoot

\end{longtable}

\endgroup

\paragraph{Overall Performance Metrics.}

The comparative evaluation shows that E-UDE delivers the strongest predictive performance among the models considered. During testing the model achieves an RMSE of 783.70, an MAE of 421.66, a MAPE of 13.36\%, and a correlation coefficient of 0.961. These metrices shows a better agreement with the real data as compared to the existing methods.

\paragraph{Train--Test Performance.}

\begin{figure}[ht!]
    \centering
    \includegraphics[width=1\linewidth]{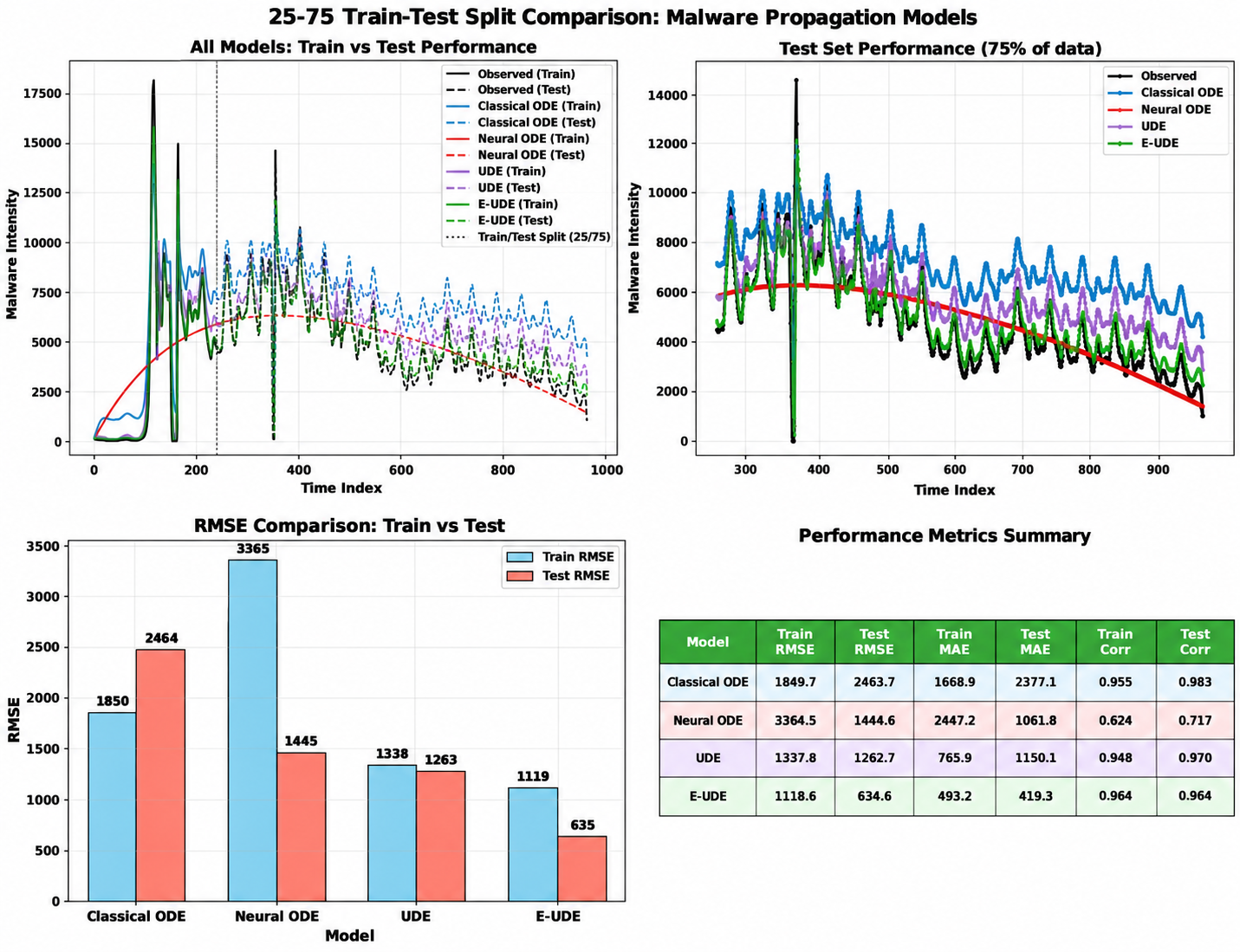}
    \caption{Comparison of prediction trajectories along with performance matrices.}
    \label{fig:prediction_trajectories_1}
\end{figure}

The predictive performance of the models were investigated using 25\% training and 75\% testing data. The results are shown in Table~\ref{tab:performance-metrics} and in Figure~\ref{fig:prediction_trajectories_1}.

The Classical ODE model exhibits the poorest predictive accuracy, obtaining a Test RMSE of 2463.66 and a Test MAE of 2377.08. These comparatively large deviations indicate that the prescribed mechanistic equations are insufficient to reproduce all of the nonlinear variations present in the observed malware propagation process. The Neural ODE achieves reduced prediction error with Test RMSE of 1444.60 and Test MAE of 1061.85. The UDE model obtains a further reduction in Test RMSE, reaching 1262.70, whereas its Test MAE increases slightly to 1150.10 compared with the Neural ODE. Finally, the proposed E-UDE gives the most accurate results over the test interval, with a Test RMSE of 634.6, a Test MAE of 419.3, and a Test Correlation of 0.964. Its substantially smaller prediction errors, combined with the high correlation coefficient, show that the estimated trajectory tracks the measured malware dynamics more closely than the other three approaches.

\begin{table*}[ht!]
\centering
\small
\caption{Comparison between the conventional Universal Differential Equation (UDE) framework and the proposed Enhanced Universal Differential Equation (E-UDE).}
\label{tab:ude_comparison}
\renewcommand{\arraystretch}{1.3}
\begin{tabular}{p{4cm}p{5.8cm}p{6.5cm}}
\hline
\textbf{Aspect} &
\textbf{Standard UDE} &
\textbf{Proposed E-UDE} \\
\hline

Model formulation &
Hybrid mechanistic model with single compartment and residual neural network. &
Hybrid cyber-mechanistic model with four detailed cybersecurity state compartments and residual neural netowrk. \\

Cybersecurity representation &
Generic compartmental dynamics. &
Models realistic enterprise cybersecurity processes, including congestion, incident response, adaptive defense, and external attack forcing. \\

Scenario analysis &
Limited capability for cybersecurity-specific scenario analysis &
Enables detailed cybersecurity-specific scenario analysis \\

Robustness to measurement noise & Moderate & High \\

Model fitting and prediction & Moderate & High \\

Residual learning &
Broad search space for neural optimization. &
Reduced search space through incorporation of mechanistic priors. \\

Train--test data splitting & Comparatively lower performance & Consistently better performance across different data splits \\

\hline
\end{tabular}
\end{table*}

\begin{table*}[ht!]
\centering
\caption{Performance metrics for all malware dynamics models.}
\begin{tabular}{lrrrr}
\toprule
Model & RMSE & MAE & MAPE (\%) & Correlation \\
\midrule
Classical ODE & 2325.87 & 2174.75 & 1503.79 & 0.948 \\
Neural ODE    & 2094.76 & 1407.10 &  532.96 & 0.664 \\
UDE           & 1281.84 & 1060.78 &   36.96 & 0.940 \\
E-UDE         & \textbf{783.70} & \textbf{421.65} & \textbf{13.36} & \textbf{0.961} \\
\bottomrule
\end{tabular}
\label{table:performance_metrics_1}
\end{table*}

\begin{table*}[ht!]
\centering
\caption{Train--test performance summary using a 25--75 split.}
\begin{tabular}{lrrrrrr}
\toprule
Model & Train RMSE & Test RMSE & Train MAE & Test MAE & Train Corr & Test Corr \\
\midrule
Classical ODE & 1849.7 & 2463.7 & 1668.9 & 2377.1 & 0.955 & 0.983 \\
Neural ODE & 3364.5 & 1444.6 & 2447.2 & 1061.8 & 0.624 & 0.717 \\
UDE & 1337.8 & 1262.7 & 765.9 & 1150.1 & 0.948 & 0.970 \\
E-UDE & 1118.6 & 634.6 & 493.2 & 419.3 & 0.964 & 0.964 \\
\bottomrule
\end{tabular}
\label{tab:performance-metrics}
\end{table*}


\paragraph{Sensitivity to Data Splitting}
\label{subsec:split_analysis}

Now we investigate the robustness of model performance under three different chronological partitions of Train--Test data, i.e., 25--75, 50--50, and 75--25. The observations were divided according to their original temporal sequence rather than being randomly assigned to the training and testing sets. Under these three data splitting setup, the four models were evaluated: the Classical ODE, Neural ODE, UDE, and the proposed E-UDE. Performance was measured using root mean squared error (RMSE), mean absolute error (MAE),and Pearson correlation coefficient (Corr). Lower RMSE and MAE indicate smaller prediction errors, whereas a correlation coefficient closer to one indicates better agreement with the temporal variation of the observed malware intensity. Detailed performance comparison for different train-test splits are available in Appendix A. Figure~\ref{fig:EUDE_TRAIN_TEST_PLOT} and Table~\ref{tab:split_comparison} summarizes the test performance of all
models across the three chronological splits.

A clear pattern can be observed across the three chronological train--test experiments. The Classical ODE consistently achieves high correlation coefficients, ranging from 0.983 to 0.988. However, its test RMSE remains relatively high, ranging from 2463.7 to 2796.8. This combination of high correlation and large prediction error suggests that the model follows the general shape of the observed trajectory but does not reproduce its absolute magnitude accurately. The Neural ODE shows a different pattern across the three splits. Its test RMSE ranges from 760.3 to 1444.6, while the corresponding correlation coefficients remain comparatively low, between 0.654 and 0.717. This indicates that lower absolute errors in some cases do not necessarily mean that the model reproduces the temporal variations of the observed data well. The results therefore suggest that RMSE alone is not sufficient for evaluating a dynamical model. The UDE model provides stronger temporal agreement than the Neural ODE, with test correlations ranging from 0.970 to 0.984. However, its test RMSE remains between 1262.7 and 1460.4. The E-UDE performs consistently well across all three train--test configurations, achieving the lowest test RMSE and MAE in each case. Its test RMSE values are 634.6, 443.3, and 493.6 for the 25--75, 50--50, and 75--25 splits, respectively. The corresponding test MAE values are 419.3, 389.5, and 442.6, while the test correlation coefficients are 0.964, 0.978, and 0.969. Its performance across the three chronological splits also remains relatively stable as the proportion of training data changes, suggesting that the enhanced formulation is less dependent on the train--test configuration.


To quantify the improvement more directly, the relative reduction in test
RMSE obtained by E-UDE can be computed as

\begin{equation}
\text{RMSE Reduction}(\%) =
\frac{\text{RMSE}_{\mathrm{baseline}}
-
\text{RMSE}_{\mathrm{E\mbox{-}UDE}}}
{\text{RMSE}_{\mathrm{baseline}}}
\times 100.
\end{equation}

Using test RMSE as the reference measure, E-UDE reduces the prediction error relative to the Classical ODE by approximately 74.4\%, 84.8\%, and 83.7\% for the 25--75, 50--50, and 75--25 splits, respectively. Compared with the standard UDE, the corresponding reductions are approximately 50.0\%, 70.4\%, and 68.7\%. The improvement is consistent across all three chronological train--test configurations rather than being limited to a single split.

\begin{table*}[ht!]
\centering
\tiny
\caption{Test-set performance across different chronological train--test splits. The best RMSE and MAE values for each split are highlighted in bold.}
\label{tab:split_comparison}
\resizebox{\textwidth}{!}{
\begin{tabular}{llccc}
\hline
\textbf{Train--Test Split} & \textbf{Model} &
\textbf{Test RMSE} & \textbf{Test MAE} & \textbf{Test Corr} \\
\hline
25--75 & Classical ODE & 2463.7 & 2377.1 & 0.983 \\
      & Neural ODE    & 1444.6 & 1061.8 & 0.717 \\
      & UDE           & 1262.7 & 1150.1 & 0.970 \\
      & E-UDE         & \textbf{634.6} & \textbf{419.3} & 0.964 \\
\hline
50--50 & Classical ODE & 2708.0 & 2692.7 & 0.988 \\
      & Neural ODE    & 1073.4 & 860.8 & 0.654 \\
      & UDE           & 1388.9 & 1361.1 & 0.984 \\
      & E-UDE         & \textbf{443.3} & \textbf{389.5} & 0.978 \\
\hline
75--25 & Classical ODE & 2796.8 & 2791.1 & 0.986 \\
      & Neural ODE    & 760.3 & 592.7 & 0.696 \\
      & UDE           & 1460.4 & 1450.0 & 0.984 \\
      & E-UDE         & \textbf{493.6} & \textbf{442.6} & 0.969 \\
\hline
\end{tabular}
}
\end{table*}

\begin{figure}[ht!]
    \centering
    \includegraphics[width=1\linewidth]{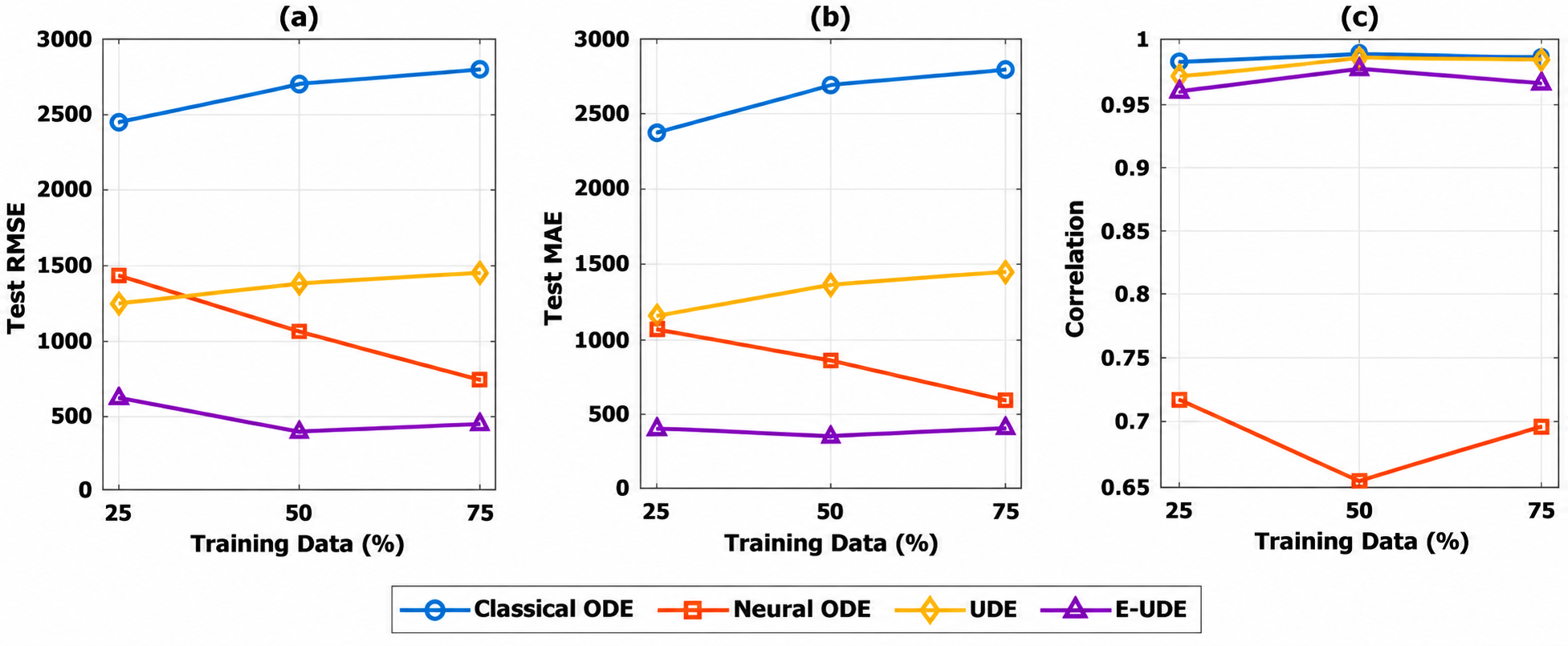}
    \caption{Test-set performance across different chronological train–test splits.}
    \label{fig:EUDE_TRAIN_TEST_PLOT}
\end{figure}

\paragraph{Extracted Symbolic Residuals}

To improve the interpretability of the proposed Enhanced Universal Differential Equation (E-UDE), the trained neural residual is analyzed after model convergence using symbolic regression. The symbolic expression is obtained only after the training process and therefore does not affect the optimization or the numerical integration of the governing differential equations. Instead, it provides an analytical approximation of the learned residual, making it possible to examine the nonlinear corrections introduced by the neural component. Let the normalized infected population be 
\[
I_{\text{norm}} = \frac{I(t)}{M_{\max}},
\]
where $I(t)$ denotes the infected population at time $t$ and $M_{\max}$ represents the maximum signal capacity. The symbolic regression procedure produces the following approximation of the learned residual for the E-UDE:

\begin{equation}
\Phi_{\text{trans}}(I_{\text{norm}}) \approx -0.1297 \left( \frac{I_{\text{norm}}}{1 + I_{\text{norm}}} \right) - 0.1297 \ln (1 + I_{\text{norm}}) - 0.1298 I_{\text{norm}} + 0.1297 I_{\text{norm}}^2.
\label{eq:symbolic}
\end{equation}

The relatively small magnitude of the coefficients in the symbolic expression indicates that the learned residual provides only limited, fine-tuned corrections to the underlying mechanistic dynamics. This behaviour is different from that of the Standard UDE, where the simpler mechanistic backbone can leave a larger portion of the observed nonlinear behaviour to the neural residual. In the E-UDE, mechanisms such as cyber defense, network effects, and congestion are already incorporated into the governing equations. As a result, the neural component does not need to compensate for the overall growth and suppression of the outbreak, but instead adjusts the dynamics where the mechanistic formulation does not fully match the observed data. Inspection of Equation~\ref{eq:symbolic} indicates that the coefficients associated with the rational, logarithmic, and linear basis functions are almost identical, each being close to $-0.1297$. In the infection-rate regime examined in this study, the functions
$I_{\text{norm}}/(1+I_{\text{norm}})$, $\ln(1+I_{\text{norm}})$, and $I_{\text{norm}}$
produce highly similar responses and exhibit a mathematical correlation exceeding $0.999$.

 The relatively small learned terms suggest that the neural network does not replace the main mechanistic processes and its role is to introduce additional corrections around the dynamics generated by the mechanistic backbone. This behavior is consistent with the intended hybrid-model design, in which the mechanistic equations describe the dominant structure of malware propagation and the neural component captures residual effects that are not adequately represented by the selected analytical terms.

To evaluate whether the symbolic representation retains the information learned by the neural network, the extracted symbolic residual was evaluated against the original neural residual generated after model training. The resulting functions exhibit close correspondence throughout the considered range. This agreement indicates that the symbolic approximation preserves the dominant behavior of the trained neural component while replacing its less transparent numerical representation with an explicit mathematical expression.

\section{Robustness Analysis}\label{robustness}

\begin{figure}[ht!]
    \centering
    \includegraphics[width=0.9\linewidth]{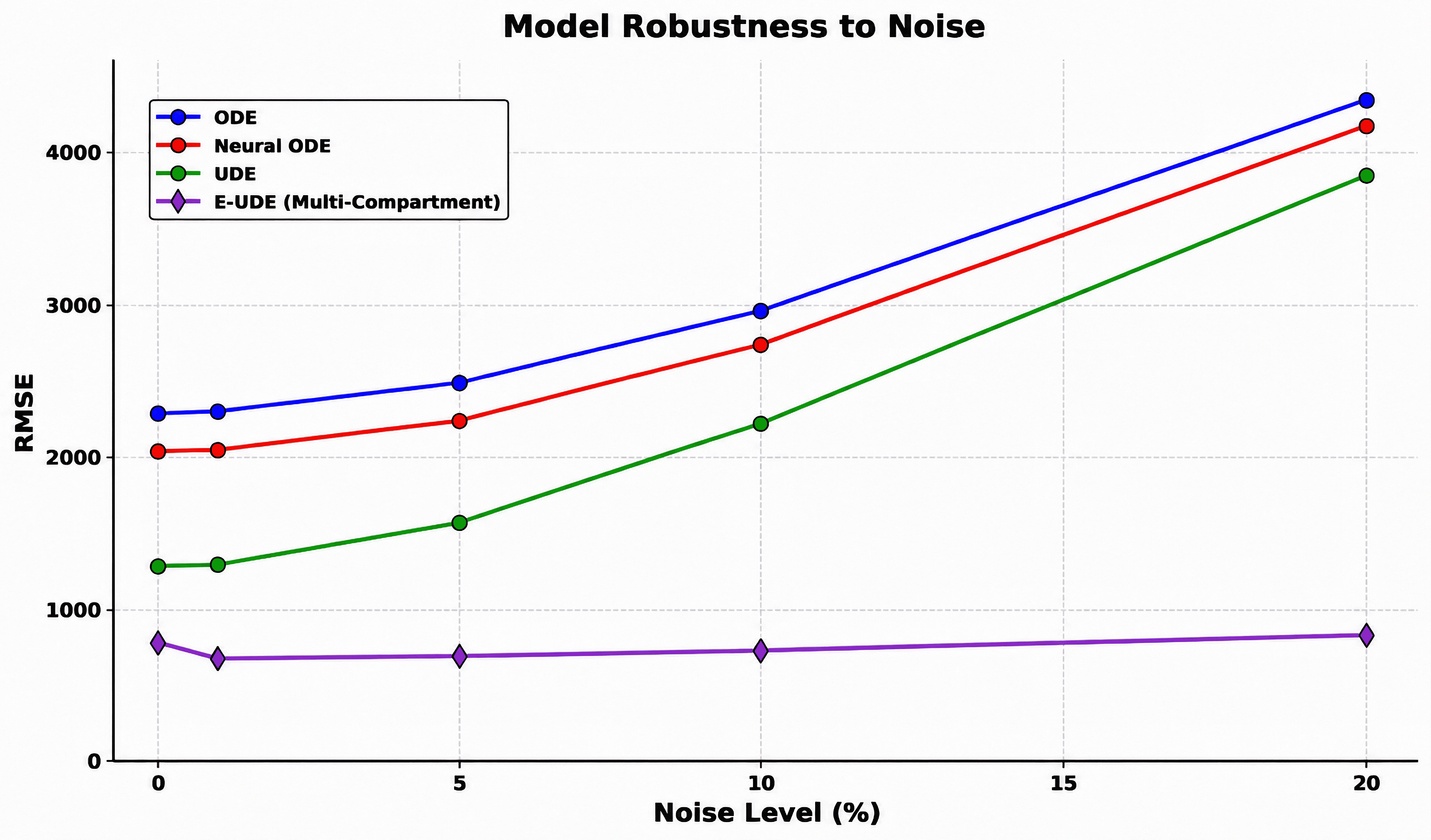}
        \caption{Comparison of model robustness with varying level of noise.}
    \label{fig:robustness}
\end{figure}

Now we investigate the robustness of the proposed E-UDE model under the data uncertainty by introducing multiple levels of noise to the observed time series data of malware propagation. The performance of the model is evaluated through the RMSE metric as a function of the noise level. The observed unperturbed dataset is denoted by a noise level of $0\%$. Then five noise strengths were introduced into the observed data with the noise levels: $0\%$, $1\%$, $5\%$, $10\%$, and $20\%$. The corresponding robustness analysis is presented in Figure~\ref{fig:robustness}.

The E-UDE shows comparatively stable predictive behavior across the tested noise range. With no added perturbation, the model obtains an RMSE of approximately $783.70$, which increases to around $830.20$ when the noise level reaches $20\%$. Although the prediction error increases as the magnitude of the perturbation grows, the overall variation remains relatively limited, which indicates that the proposed formulation retains a substantial degree of predictive stability even when the input observations become increasingly noisy. This suggests that the E-UDE can preserve much of its predictive accuracy even when the observed malware propagation data are affected by measurement or observational noise. The difference becomes clearer when the E-UDE is compared with the baseline models. Over the same increase from 0\% to 20\% noise, the RMSE rises by approximately 2069.2 for the Classical ODE, 2150.9 for the Neural ODE, and 2574.4 for the Standard UDE. In comparison, the E-UDE shows only a 43.27-point increase. The much smaller change indicates that the proposed model is less sensitive to the perturbations introduced into the data.Overall, the noise experiment indicates that the E-UDE retains comparatively stable predictive performance across the tested noise levels. Its lower increase in RMSE, compared with the baseline models, supports the robustness of the proposed formulation under increasingly noisy observations.

\section{Scenario analysis}\label{scenarios}
In this section, we perform scenario analysis using the E-UDE framework. By modifying cyber-epidemiological conditions or operational defense parameters within the trained E-UDE model, a wide range of hypothetical scenarios can be systematically investigated. This enables us to evaluate the impact of different intervention measures to mitigate malware propagation.

\subsection{Impact of External Attack Forcing Intensity}

To evaluate the sensitivity of the proposed E-UDE framework to external cyber threats, we performed an operational what-if scenario analysis by varying the external attack forcing signal $F(t)$. This forcing term represents attacks originating outside the enterprise network, including automated vulnerability scanning, phishing campaigns, botnet reconnaissance, and zero-day exploitation attempts. Unlike the internal malware transmission process, which depends on infected hosts already present in the network, $F(t)$ continuously introduces new infection pressure from external attackers. The purpose of this scenario analysis is to quantify the effect of externally generated attack pressure on malware propagation while holding the internal defensive parameters fixed. In this way, changes in the resulting infection trajectories can be attributed primarily to variations in the external forcing signal.

The calibrated attack signal, $F_{0}(t)$, was multiplied by five scaling coefficients: $0.5\times$, $1.0\times$, $2.0\times$, $5.0\times$, and $10.0\times$. The case $0.5\times F_{0}(t)$ represents a situation in which perimeter security controls substantially suppress incoming malicious activity. In contrast, the larger multipliers correspond to progressively more severe external attack conditions, representing scenarios ranging from increased background scanning to highly coordinated and intensive exploitation campaigns.

\begin{figure}[ht!]
    \centering
    \includegraphics[width=1\linewidth]{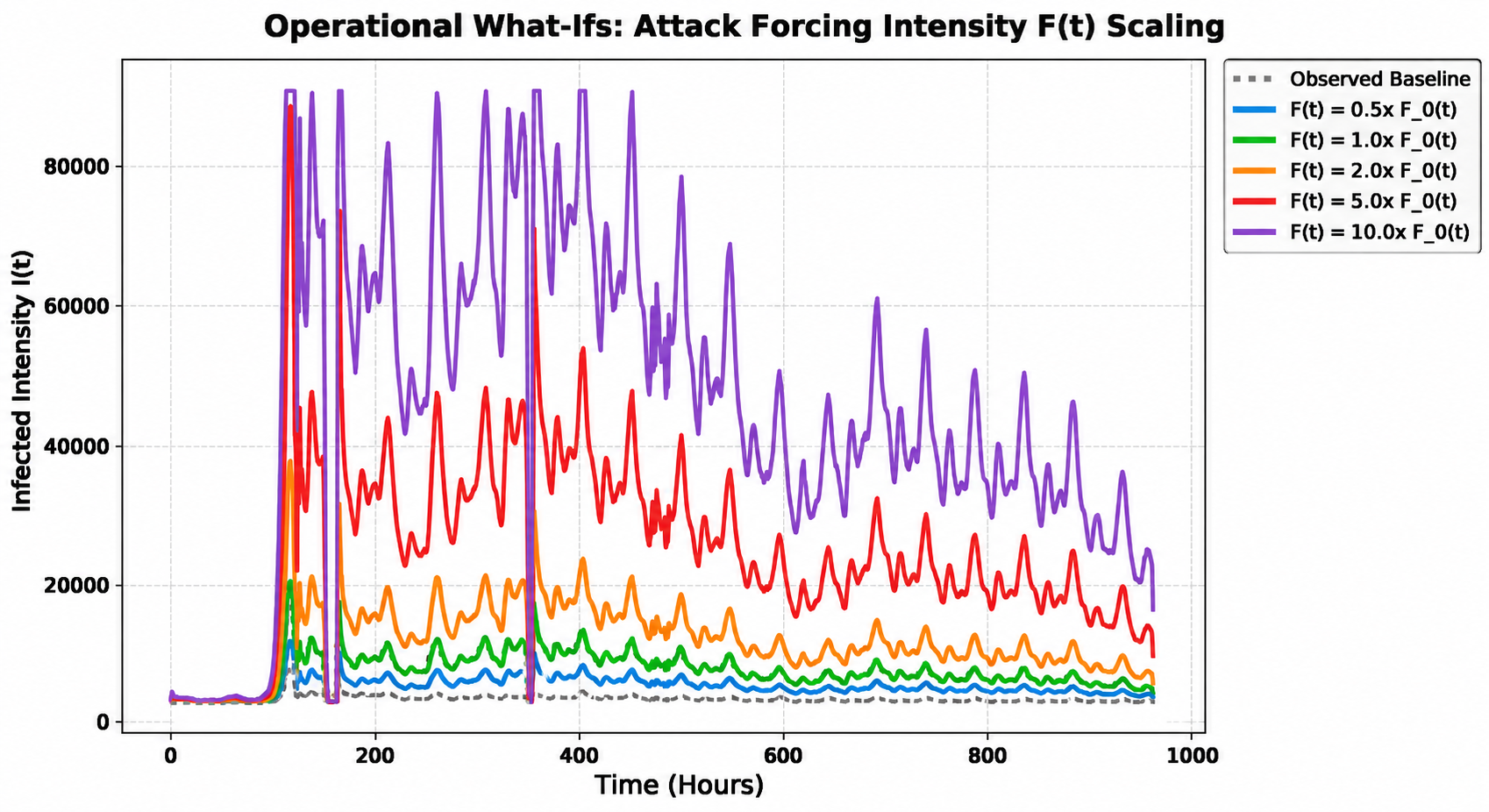}
    \caption{Attack Forcing Intensity Effects on Malware Propagation}
    \label{fig:impact_attack_forcing}
\end{figure}

The resulting trajectories are presented in Figure~\ref{fig:impact_attack_forcing}. A clear dependence of malware propagation on the magnitude of external forcing is observed. When the attack signal is reduced to $0.5\times F_{0}(t)$, both the maximum infection level and the total outbreak burden decline considerably. This behavior indicates that effective filtering of malicious traffic at the network boundary can limit the number of attacks reaching internal systems and thereby reduce subsequent propagation. For the reference case, $1.0\times F_{0}(t)$, the simulated trajectory corresponds to the baseline scenario calibrated against the historical propagation data.

Increasing the forcing to $2.0\times F_{0}(t)$ produces a substantial increase in both the peak number of infections and the accumulated outbreak size, with both quantities exhibiting approximately a twofold increase relative to the lower-intensity case. A different behavior emerges at $5.0\times F_{0}(t)$ and $10.0\times F_{0}(t)$, where the system approaches a saturation region. Under these extreme attack conditions, a large fraction of the enterprise population becomes infected over a relatively short interval, and the infection trajectory approaches the maximum host capacity imposed by the model. Once the susceptible population is nearly depleted, additional increases in external forcing have a progressively smaller effect because there are fewer remaining hosts available for infection.

Overall, the experiment identifies external attack intensity as a major factor governing the scale and speed of malware propagation. The findings also emphasize the value of preventive controls deployed at the network perimeter, including firewalls, intrusion prevention systems, email-security filtering, and threat-intelligence mechanisms. Suppressing malicious activity before it reaches vulnerable internal hosts can reduce both the maximum infection level and the cumulative outbreak impact without requiring changes to the internal defensive dynamics of the model.

\subsection{Impact of Network Topology}
Although the E-UDE model successfully combines mechanistic epidemiological modeling with neural residual learning, it assumes that malware transmission depends primarily on homogeneous compartment populations such as susceptible, infected, response, and congestion. However, modern enterprise networks exhibit significantly heterogeneous graph structures where the propagation behavior is strongly influenced by the underlying communication topology. Two networks containing the same number of infected devices may experience entirely different propagation dynamics because their connectivity patterns, centrality distributions, and clustering characteristics are fundamentally different.
\subsubsection{Topology-Aware Enhanced Universal Differential Equation}

To study the implicit impact of network topology we further extend the proposed E-UDE by explicitly incorporating dynamic network topology into both the mechanistic transmission model and the neural residual component. 

\paragraph{Temporal Network Topology Representation.}

At every simulation time step, the malware propagation graph is analyzed to extract a compact representation of the evolving network structure. Instead of using the raw adjacency matrix, five graph-theoretic descriptors are considered,

\begin{equation}
T(t)=
\left[
T_{\mathrm{degree}},
T_{\mathrm{betweenness}},
T_{\mathrm{clustering}},
T_{\mathrm{entropy}},
T_{\mathrm{density}}
\right],
\label{eq:topology_vector}
\end{equation}
where each component captures a different structural characteristic of the infected communication graph. The average degree measures the overall network connectivity and reflects the number of communication links available for malware propagation. Betweenness centrality captures the role of bridge-like devices that can connect otherwise separate parts of the network and facilitate the spread of malware across them. The clustering coefficient reflects how strongly a device's neighboring nodes are connected to one another, which can influence the speed of local propagation. Graph entropy provides a measure of the structural complexity of the communication network, while graph density indicates how many of the possible communication links are actually present. These network descriptors are used only as external inputs to the model. They are not treated as additional compartments or dynamic state variables. Instead, their time-dependent values are supplied to the model to adjust the transmission process and the behavior of the neural component. They are therefore not governed by separate differential equations and do not affect the population balance $S+I=N$.

\paragraph{Topology-Aware Transmission.}
In the original E-UDE formulation, malware transmission depends primarily on the epidemiological states of the system. This can be restrictive for enterprise networks because their communication structure can strongly influence how quickly an infection spreads. For example, a network with many interconnected devices provides more opportunities for lateral movement than a sparse network, even when the two networks have the same number of infected devices. To account for this effect, the proposed formulation incorporates network-topology information into the transmission process.

To account for this phenomenon, the physical transmission rate is multiplied by a topology scaling function,

\begin{equation}
\Phi_{\mathrm{topo}}(T)
=
\max
\left(
0.05,
\min
\left(
5.0,
\frac{1}{2}
\left(
\frac{T_{\mathrm{degree}}}{0.272}
+
\frac{T_{\mathrm{clustering}}}{0.535}
\right)
\right)
\right).
\label{eq:topology_scale}
\end{equation}

The empirical constants 0.272 and 0.535 correspond to the average normalized degree and clustering coefficient observed within the malware dataset. The lower and upper bounds prevent unrealistic amplification of the transmission rate while allowing the physical model to adapt continuously to changes in network connectivity.

\paragraph{Topology-Aware Neural Residual.}

The neural residual is expanded to jointly learn from both epidemiological states and network topology. Instead of using only $
[S,I,R,C],
$

the neural network now receives

\begin{equation}
\mathbf{x}
=
\left[
\frac{S}{N},
\frac{I}{N},
\frac{R}{N},
\frac{C}{N},
T_{\mathrm{degree}},
T_{\mathrm{betweenness}},
T_{\mathrm{clustering}},
T_{\mathrm{entropy}},
T_{\mathrm{density}}
\right].
\label{eq:topology_input}
\end{equation}

The topology-aware neural residual is therefore defined as

\begin{equation}
w_{\mathrm{res}}\phi_{\mathrm{trans}}(S,I,R,C,T)
=
\lambda_{\mathrm{res}}
\tanh
\left(
\frac{
NN_{\theta}(\mathbf{x})
}
{\lambda_{\mathrm{res}}}
\right),
\label{eq:topology_residual}
\end{equation}

where \(NN_{\theta}\) denotes the multilayer perceptron and \(\lambda_{\mathrm{res}}\) represents the residual bound.

\paragraph{Modified Topology-Aware Governing Equations.}

Integrating the topology scaling function and topology-aware neural residual yields the following governing equations.

The susceptible compartment becomes

\begin{equation}\label{modified_S}
\frac{dS}{dt}
=
-
\left[
\beta(t)
\frac{I}{N}
e^{-\alpha C/N}
\,
\Phi_{\mathrm{topo}}(T)
+
\frac{F(t)}{N}
+
w_{\mathrm{res}}\phi_{\mathrm{trans}}(S,I,R,C,T)
\right]
S
+(\gamma+\delta R+\mu C)I - f_{def} F(t) \frac{S}{N},
\end{equation}

and the infected compartment is

\begin{equation}\label{modified_I}
\frac{dI}{dt}
=
\left[
\beta(t)
\frac{I}{N}
e^{-\alpha C/N}
\,
\Phi_{\mathrm{topo}}(T)
+
\frac{F(t)}{N}
+
w_{\mathrm{res}}\phi_{\mathrm{trans}}(S,I,R,C,T)
\right]
S
-
(\gamma+\delta R+\mu C)I.
\end{equation}

The equations for $R$ and $C$ are same as given in equations \ref{eq:response} and \ref{eq:congestion} respectively. Only the transmission dynamics are modified by topology, preserving the physical interpretation of the response and congestion compartments. $f_{\mathrm{def}}$ is the proactive defense rate, and the last term
$-f_{\mathrm{def}}F(t)S/N$ represents pre-emptive hardening of healthy hosts upon detecting external attacks $F(t)$. Unlike reactive terms that clean already-infected nodes, this shields susceptible hosts before compromise. With it, early warnings trigger automated defenses to curb outbreaks; without it, external traffic hits unprotected hosts, artificially inflating infection peaks.









\subsubsection{Network Topology-Based Scenarios}

\begin{figure}[ht!]
    \centering
    \includegraphics[width=1\linewidth]{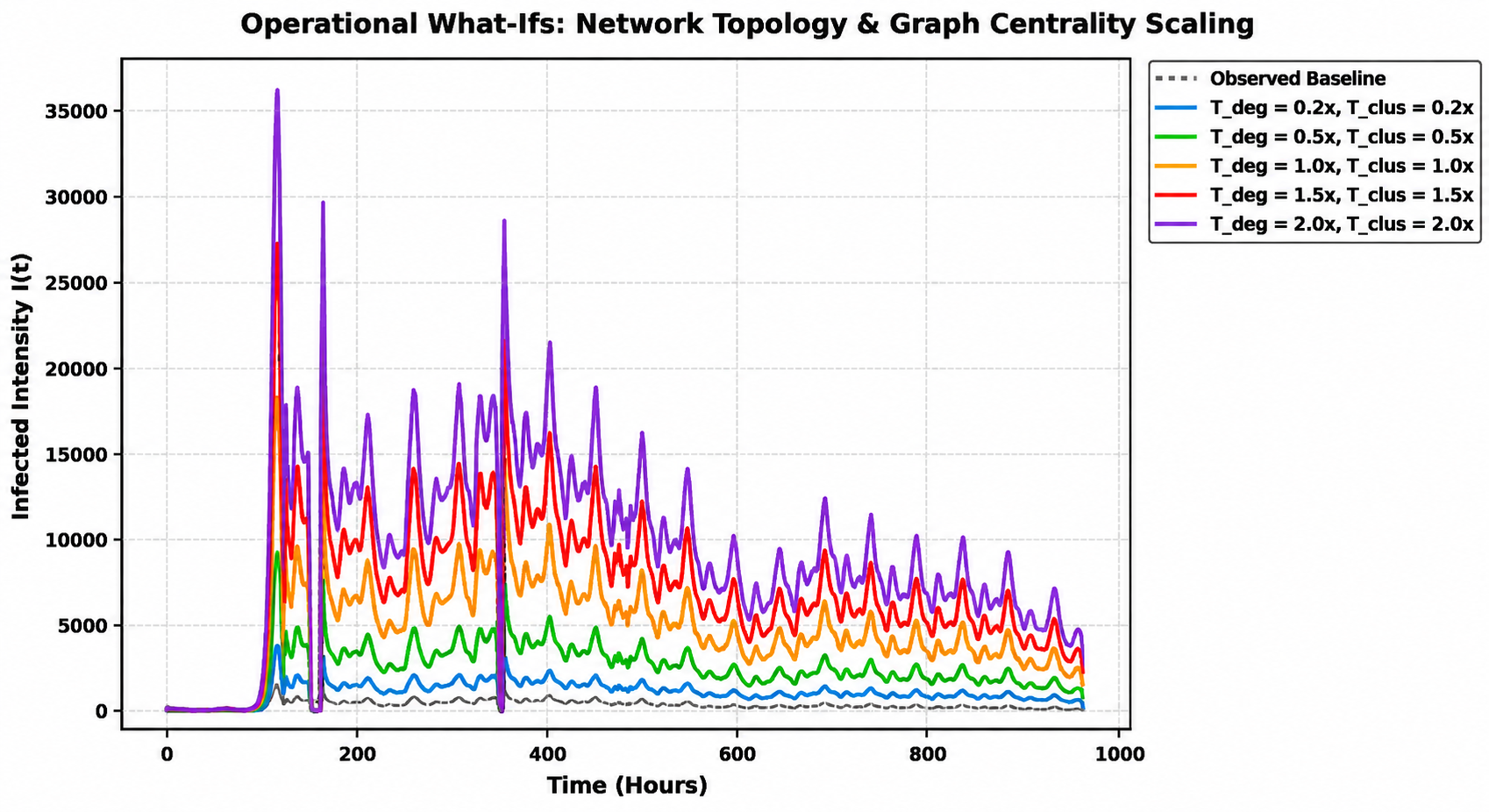}
    \caption{Impact of network topology characteristics on the infection progression.}
    \label{fig:topology_scenario}
\end{figure}

We now perform different scenarios to investigate the impact of network topology on infection progression. For this purpose, we simulate the modified model equations \eqref{modified_S}, \eqref{modified_I}, \ref{eq:response} and \ref{eq:congestion} with different choices of $T_{\mathrm{degree}}$ and $T_{\mathrm{clustering}}$ in the formula for $\Phi_{\mathrm{topo}}(T)$ given in Eq.~\eqref{eq:topology_scale}. The scenario results are shown in Figure~\ref{fig:topology_scenario}. As shown in Figure~\ref{fig:topology_scenario}, increasing $T_{\mathrm{degree}}$ and $T_{\mathrm{clustering}}$ leads to higher infection counts throughout the simulation. An increase in the average node degree means that each device has more communication links through which malware can potentially reach other devices. In the same way, a higher clustering coefficient indicates stronger connections among neighboring devices, allowing malware to spread through closely connected groups of devices and increasing the number of opportunities for local transmission. Both quantities contribute to the topology scaling factor $\Phi_{\mathrm{topo}}(T)$. As this factor increases, the effective transmission rate also increases, leading to faster growth in the infected population and a higher total number of infected devices over the simulation period. The results therefore suggest that network topology plays an important role in malware propagation, with highly connected and strongly clustered networks providing more opportunities for infections to spread.

\subsection{Impact of several cybersecurity scenarios}

Now we illustrate the temporal evolution of the four state variables of the E-UDE model, namely \(S(t)\), \(I(t)\), \(R(t)\), and\(C(t)\), under five chosen cybersecurity scenarios as follows:

\begin{itemize}
    \item \textbf{Scenario-0 (Baseline):} Reference scenario with trained parameter values.
    
    \item \textbf{Scenario-1 (EDR Boost):} Enhanced detection, quarantine, and remediation with accelerated defensive isolation and recovery (\(\delta = 15.0\times,\ \rho = 1.2\times,\ f_{\text{def}} = 1.8\times\)).
    
    \item \textbf{Scenario-2 (SOC Saturation):} Quadrupled containment decay, diminished remediation speed, and impaired detection (\(\omega = 4.0\times,\ \delta = 0.4\times,\ \gamma = 0.625\times,\ f_{\text{def}} = 0.5\times\)).
    
    \item \textbf{Scenario-3 (Overload):} Excessive background malicious traffic injection and network bandwidth exhaustion with elevated congestion coefficients (\(\sigma = 3.0\times,\ \alpha = 2.5\times\)).
    
    \item \textbf{Scenario-4 (OOD Shock):} Initial perturbation with a five-fold increase in initial infected hosts (\(I(0) = 5I_0\)).
\end{itemize}

\begin{figure}[ht!]
    \centering
    \includegraphics[width=1\linewidth]{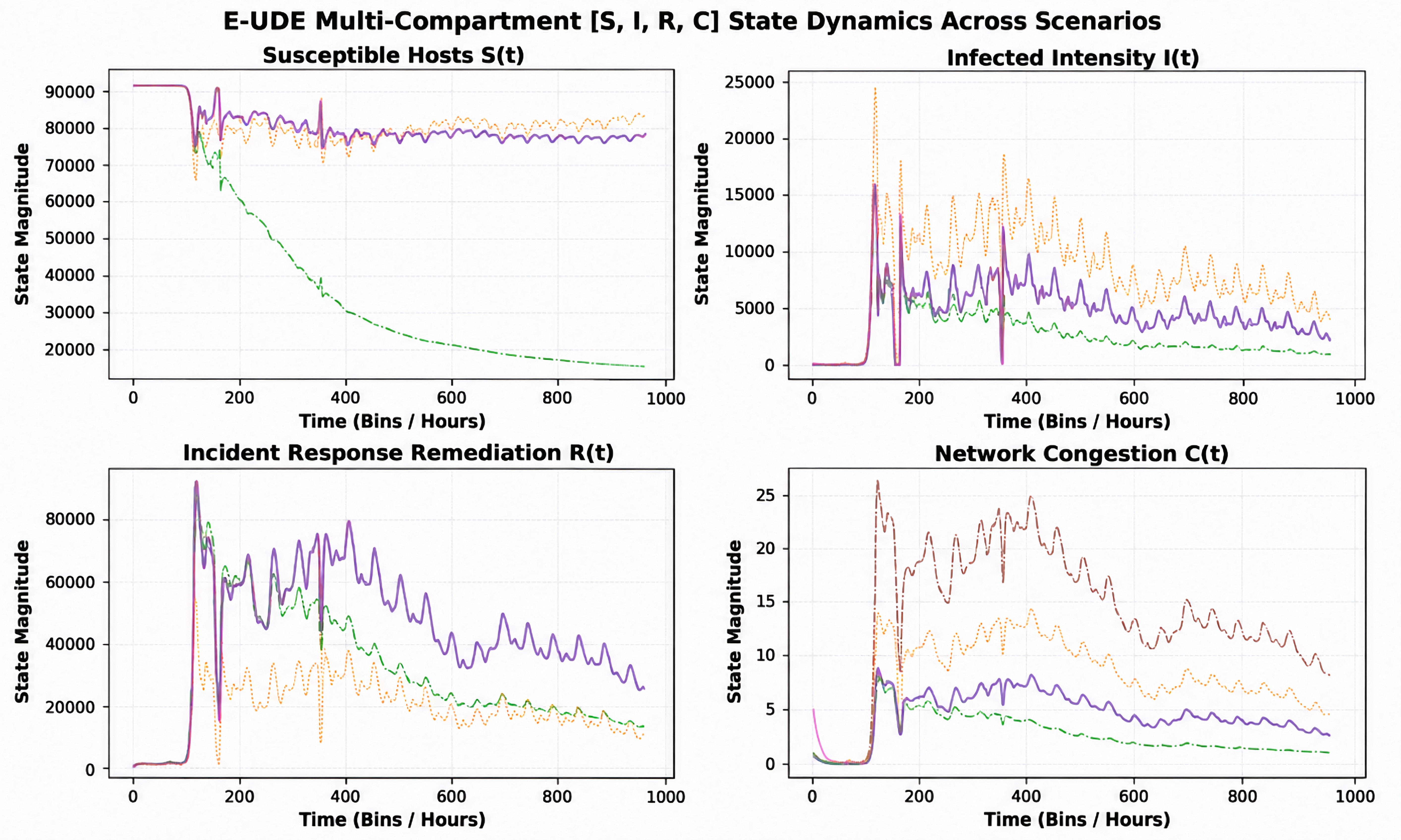}
    \caption{Multi-compartmental state dynamics of E-UDE model under different cybersecurity scenarios.}
    \label{fig:state_dynamics_scenarios}
\end{figure}

The results are shown in Figure~\ref{fig:state_dynamics_scenarios}. For \(S(t)\), the baseline, overload, and OOD shock cases show very similar trajectories, with the solutions settling around \(7.87\times10^4\)--\(7.88\times10^4\). The mean value is approximately \(8.09\times10^4\). In contrast, the SOC saturation case reaches a higher steady level of about \(8.41\times10^4\), which can be attributed to the reduced host recovery and clearance capacity under sustained SOC load. In contrast, the EDR boost drives \(S(t)\) to decrease steadily through aggressive proactive quarantine, reaching approximately \(1.55\times10^4\) (minimum \(1.54\times10^4\)).

For \(I(t)\), SOC saturation generates the largest transient peak (\(24,492 \approx 2.45\times10^4\), a \(+54.3\%\) surge relative to baseline) as well as the highest sustained infection burden (final infection of \(4,018 \approx 4.02\times10^3\) and \(+73.9\%\) cumulative infection volume). Conversely, the EDR boost drastically improves epidemic containment, dampening peak infection to \(1.49\times10^4\) (\(-6.0\%\)), slashing cumulative infections by \(41.9\%\) (\(2.65\times10^6\) versus \(4.56\times10^6\)), and cutting final infected hosts by more than half down to \(1,097 \approx 1.10\times10^3\) (\(-51.8\%\)). Both overload and OOD shock rapidly converge back to the nominal baseline trajectory (\(\approx 2.27\times10^3\) final infected hosts).

For \(R(t)\), the nominal baseline, EDR boost, overload, and OOD shock all exhibit a vigorous initial incident response surge peaking at \(91,440 \approx 9.14\times10^4\). In contrast, SOC saturation severely cripples response capabilities, limiting the peak remediation level to only \(54,576 \approx 5.46\times10^4\) (a \(40.3\%\) degradation) and maintaining the weakest post-outbreak remediation level throughout the observation horizon (final \(R(t) \approx 1.06\times10^4\) versus baseline \(2.56\times10^4\)).

For \(C(t)\), the overload case shows the largest congestion peak, reaching approximately \(26.9\), with a mean value of \(14.4\) and a final value of \(8.63\). SOC saturation also results in elevated congestion, with a peak of about \(14.6\) and a final value of \(4.74\), reflecting the effect of sustained network activity. In contrast, the EDR boost keeps congestion comparatively low, with a peak of \(8.48\) and a final value of \(1.24\), compared with \(2.87\) for the baseline case.

Across the four experimental conditions, the system states exhibit distinct transient behaviors; nevertheless, each trajectory remains bounded and ultimately settles toward a stable regime. Under SOC saturation, the infection persists for a longer duration and the overall network burden increases, which can be attributed primarily to the reduction in the effective capacity of the security analysts. The DDoS overload condition predominantly influences network-level dynamics, producing larger communication delays and increased packet loss. In contrast, enhancing the EDR capability strengthens the automated containment mechanism, thereby decreasing both the infection magnitude and the associated network congestion. The OOD shock also does not lead to instability or unbounded evolution of the state variables. Although the system initially experiences a five-fold increase in the external disturbance, its trajectories progressively move back toward the bounded operating regime. These observations suggest that the learned E-UDE model maintains stable behavior and exhibits reasonable robustness when exposed to operating conditions that are substantially different from those represented in its training data.

\section{Model Performance on the Witty Worm Dataset}
\label{cross_data}

Furthermore, the E-UDE model framework was trained exclusively on the Witty Worm dataset to check the model performance on different type of malware propagation dataset. The initial infected state, $I_0$, was obtained from the first observation of the three-point smoothed Witty Worm intensity time series and was lower-bounded by one. The total modeled population, $N$, was assumed to be larger than the maximum observed infection intensity in the Witty Worm dataset. The initial susceptible population was assumed to be $
S_0 = N-I_0.$ The trained model parameter values are summarized in Table~\ref{tab:witty_worm_parameters}.

\begin{figure}[ht!]
    \centering
    \includegraphics[width=1\linewidth]{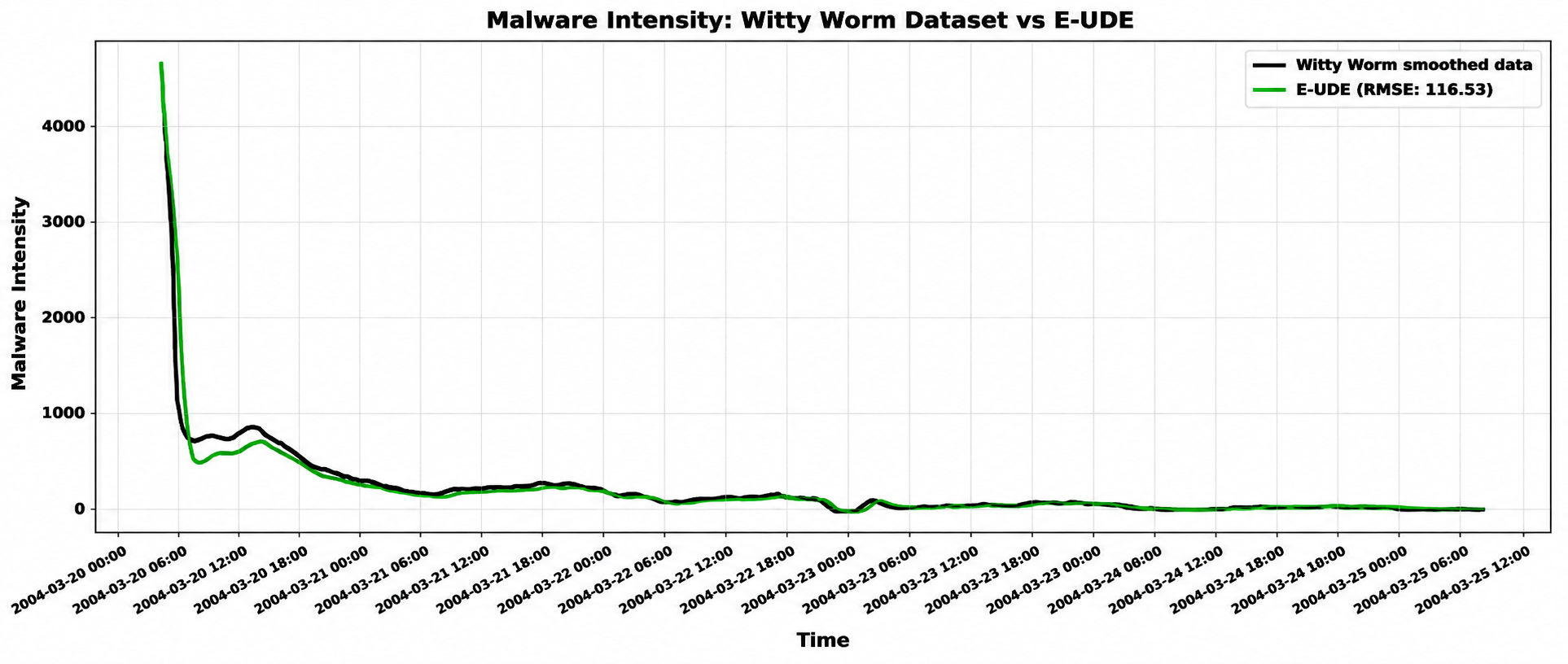}
    \caption{E-UDE prediction compared with the observed Witty Worm
    infection-intensity time series.}
    \label{fig:witty_comparison}
\end{figure}

Figure~\ref{fig:witty_comparison} compares the observed Witty Worm outbreak with the trajectory predicted by the trained E-UDE model. 
\begin{table}[ht!]
\centering
\caption{Initial and Optimized E-UDE Parameters on the Witty Worm Dataset}
\label{tab:witty_worm_parameters}
\small
\begin{tabular}{l c c c}
\hline
\textbf{Parameter} & \textbf{Unit} & \textbf{Initial Value} & \textbf{Optimized Value} \\
\hline
$\beta_0$ & [--] & 0.00600001 & 0.00581285 \\
$\gamma$ & [day$^{-1}$] & 0.950000 & 0.980295 \\
$\delta$ & [host$^{-1}$ day$^{-1}$] & $2.50100\times10^{-7}$ & $2.58119\times10^{-7}$ \\
$\mu$ & [host$^{-1}$ day$^{-1}$] & $5.00010\times10^{-6}$ & $5.14953\times10^{-6}$ \\
$\rho$ & [day$^{-1}$] & 1.05000 & 1.08358 \\
$\omega$ & [day$^{-1}$] & 0.150000 & 0.145403 \\
$\sigma$ & [day$^{-1}$] & 0.00030001 & 0.000290754 \\
$\tau$ & [day$^{-1}$] & 0.060000 & 0.0581388 \\
$p_{\mathrm{decay}}$ & [day$^{-1}$] & 1.25000 & 1.28895 \\
$\alpha$ & [--] & 0.180000 & 0.185570 \\
$\xi$ & [--] & 0.030000 & 0.0290901 \\
$w_1$ (memory) & [--] & 0.600000 & 0.630213 \\
$w_2$ (memory) & [--] & 0.300000 & 0.277314 \\
$w_3$ (memory) & [--] & 0.100000 & 0.0924729 \\
Neural residual weight & [--] & 0.400000 & 0.383699 \\
Defense forcing & [day$^{-1}$] & 0.010000 & 0.010815 \\
Neural residual bound & [--] & 2.50100 & 2.42466 \\
\hline
\end{tabular}
\end{table}
The prediction error is calculated as RMSE = 116.53. This shows a close agreement between the predicted and observed infection-intensity trajectories.

\section{Discussion and Conclusion}\label{discussion}

This work introduces an Enhanced Universal Differential Equation (E-UDE) framework that combines detailed mechanistic cyber-epidemiological modeling with residual neural learning for malware propagation. The proposed framework explicitly incorporates cybersecurity mechanisms, including malware transmission, network topology, congestion, external attacks, defensive actions, and security degradation, into the governing equations. This multi-compartmental mechanistic model enables us to improve the model predictions and reduces the contributions from the neural residual counterpart.

This detailed mechanistic structure reduces the burden on the neural network to learn known system dynamics and consequently narrows the residual learning space. As a result, the E-UDE demonstrates improved model fitting and predictive performance, greater robustness to measurement noise, and more consistent performance across different train--test data partitions. 

Furthermore, unlike the conventional UDE framework, the proposed formulation enables detailed scenario-based analysis of cybersecurity conditions, thereby providing a more interpretable and flexible framework for investigating the effects of different attack, defense, and network conditions. The topology-based simulations demonstrate that increasing the average node degree or clustering coefficient consistently produces larger malware outbreaks. The cybersecurity scenario analysis shows that the model outputs are different due to the changes in defensive capacity and network conditions. In particular, EDR enhancement reduces the infection burden, while SOC saturation and overload lead to higher infection or congestion levels. Even under the five-fold OOD shock, the trajectories remain bounded and return toward the usual regime, suggesting that the E-UDE model can handle substantial changes in the system conditions without losing stability.

Moreover, our E-UDE model is able to perform accurately in different types of malware propagation dataset. To examine this we employed thye model to two different types of malware dataset, namely, the Code Red dataset and Witty Worm dataset. The Code Red primarily targeted Microsoft IIS web servers and propagated through multiple waves over several days, whereas Witty Worm exploited vulnerabilities in Internet Security Systems (ISS) firewall products and spread rapidly through high-rate UDP scanning. The E-UDE model is able to learn and predict with reasonably good accuracy for both the dataset. Further evaluation on additional malware datasets and network environments would be required to establish broader applicability of the E-UDE model.

Despite its advantages, the present framework has some scopes open for further modification and refinement. The model considers aggregate population dynamics and assumes fixed network topology during each simulation. The current framework does not explicitly account for changes in network structure over time, differences in device characteristics, changes in attacker behavior, or the stochastic nature of malware transmission. Future investigations may incorporate these aspects together with online learning mechanisms that continuously update the model as new observations become available. Integrating graph-based representations could further improve the characterization of complex and dynamically changing enterprise environments. Such extensions would allow the E-UDE framework to capture structural and behavioral changes that are difficult to represent using a fixed system formulation.

A further research direction is the integration of the E-UDE model with optimization and optimal-control techniques. In this setting, the learned dynamical system could serve as a basis for determining effective combinations of defensive actions during a malware outbreak. For instance, the framework could be used to optimize how available resources are distributed among vulnerability patching, host isolation, network-traffic filtering, and enhanced security monitoring. The corresponding control objective would be to suppress the spread of malware while simultaneously limiting the operational expenditure associated with these interventions. This would broaden the application of the E-UDE framework beyond modeling and forecasting system behavior, enabling it to function as a quantitative tool for cybersecurity resource allocation and response planning.

\section*{Code Availability}

 The associated codes are available at:
\url{https://github.com/Vineet-Kumar-001/E-UDE-Model-Research.git}

\section*{Data Availability}

All data used in this study are available from the sources cited in the manuscript.

 \section*{Acknowledgement}
 This research is partly funded by Ministry of Electronics and Information Technology (MeitY), Government of India, under the Cybersecurity Research \& Development Scheme.

\section*{Conflict of Interest}
The authors declare that they have no conflict of interest.


\section{Appendix A}

\paragraph{Temporal Aggregation.}

Using the preprocessing technique described in \cite{pappu2025understanding}, we transformed the raw dataset into conventional date--time representations. Subsequently, the individual scan records were mapped to consecutive 30-minute windows, with the total number of scan events within each window computed to quantify malware activity during that period. This aggregation process defines the infection intensity associated with each sampling interval. Consequently, the original observations, which occur at nonuniform time instants, were transformed into a regularly sampled temporal sequence that captures the evolution of malware activity over time.

\paragraph{Smoothing.}

The scan data contain short-term variations that can arise from measurement noise and fluctuations in network activity. To reduce these fluctuations and obtain a more stable signal for model training and numerical integration, a three-point moving average was applied to the infection intensity series. The smoothed value $\tilde{I}_i$ is defined as

\begin{equation} \tilde{I}_i= \begin{cases} \dfrac{I_1+I_2}{2}, & i=1, \\[6pt] \dfrac{I_{i-1}+I_i+I_{i+1}}{3}, & 1<i<n, \\[6pt] \dfrac{I_{n-1}+I_n}{2}, & i=n. \end{cases} \end{equation}

For the points at the beginning and end of the series, the average is calculated using the available neighboring values. This smoothing step reduces rapid fluctuations in the observations while retaining the main trend of the malware outbreak, providing a cleaner signal for the following modeling stages.

\paragraph{Continuous-Time Interpolation.}

To enable continuous-time simulation, the smoothed infection intensity values were transformed into a continuous function using linear interpolation. For two consecutive time points $t_i$ and $t_{i+1}$, the infection intensity function $\eta(t)$ is defined as

\begin{equation}
\eta(t)=
\tilde{I}_i
+
\frac{\tilde{I}_{i+1}-\tilde{I}_i}
     {t_{i+1}-t_i}
(t-t_i),
\qquad
t_i \leq t \leq t_{i+1}.
\end{equation}


\subsubsection{50--50 Train--Test Split}

\begin{figure*}[!htbp]
    \centering
    \includegraphics[width=1\textwidth]{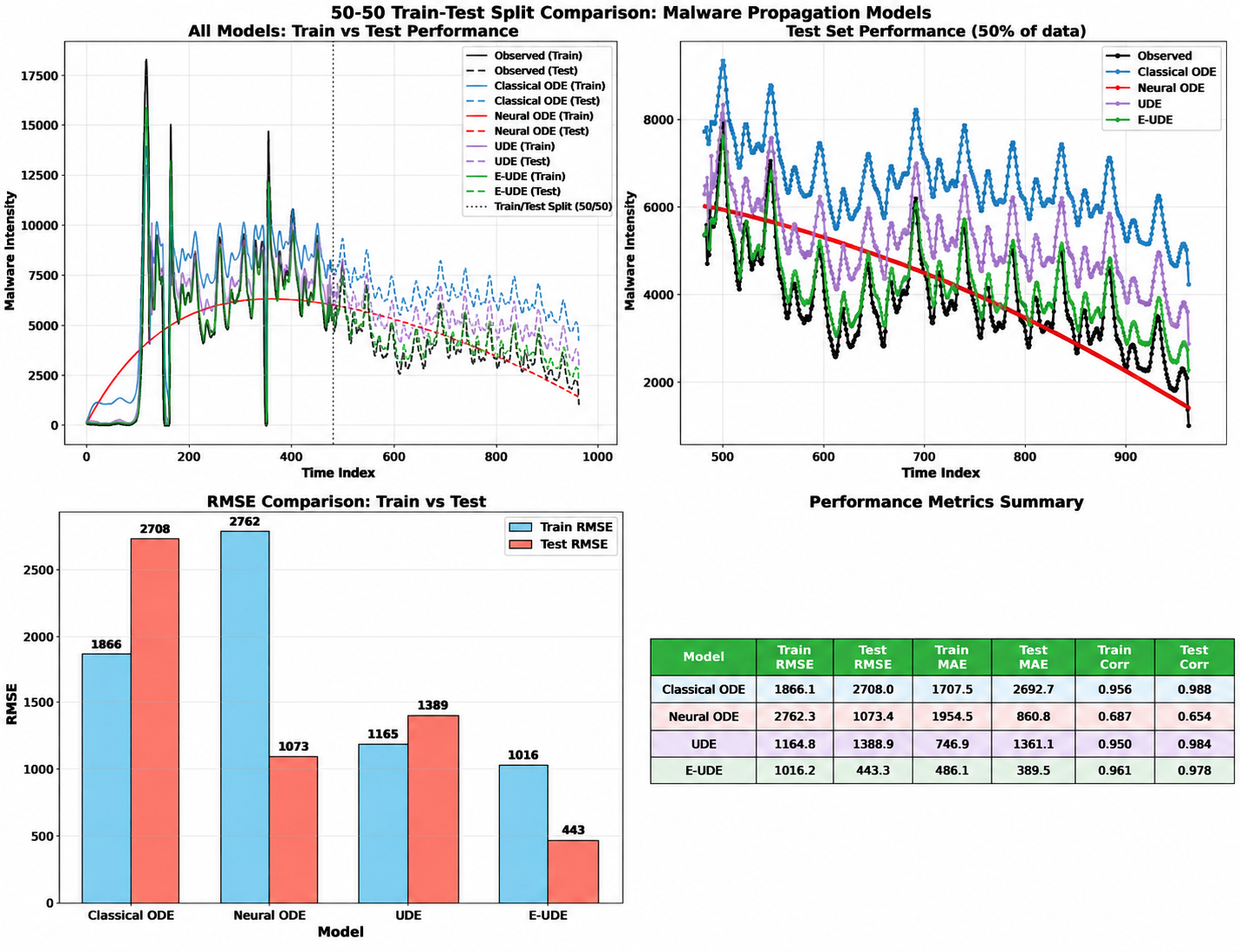}
        \caption{Performance comparison of the Classical ODE, Neural ODE, UDE,
    and E-UDE models using a 50--50 chronological train--test split.}
    \label{fig:split5050}
\end{figure*}

The second experiment uses an equal division of the sequence, with 50\%
of the observations used for training and the remaining 50\% for testing.
The corresponding results are presented in
Figure~\ref{fig:split5050}.

\subsubsection{75--25 Train--Test Split}

\begin{figure*}[!htbp]
\centering
\includegraphics[width=1\textwidth]{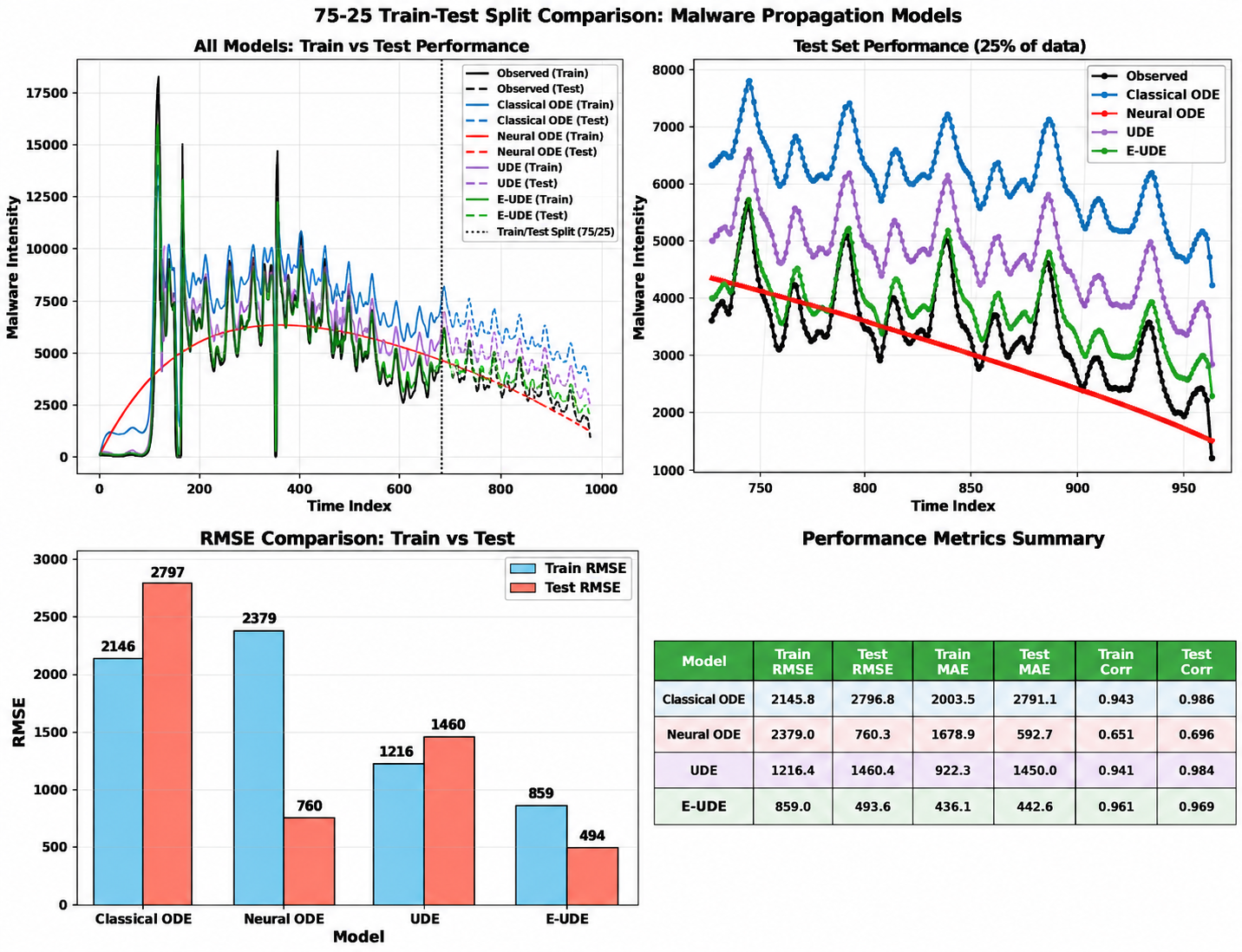}
\caption{Performance comparison of the Classical ODE, Neural ODE, UDE,
and E-UDE models using a 75--25 chronological train--test split.}
\label{fig:split7525}
\end{figure*}

This experiment uses 75\% of the observations, in chronological order, for training and the remaining 25\% for testing. The corresponding results are presented in Figure~\ref{fig:split7525}.

\end{document}